\documentclass[pre,aps,showpacs,preprint,superscriptaddress]{revtex4}
\usepackage{graphicx}
\usepackage{amsmath}
\usepackage{amssymb}

\newcommand{\e}{\mbox{\large\em e}}
\begin{document}

\title[Generalized Dicke and Jahn-Teller model]{Quantum phase transitions and quantum chaos in generalized
Dicke and Jahn-Teller polaron model and finite-size effects}

\author{Eva Majern\'{\i}kov\'a}
\email{eva.majernikova@savba.sk}

\affiliation{Institute of Physics, Slovak Academy of
Sciences, D\'ubravsk\'a cesta 9,
SK-84 511 Bratislava, Slovak Republic}

\author{Serge Shpyrko}
\email[e-mail:]{serge_shp(at)yahoo.com}
 \affiliation{Institute for Nuclear Research,
Ukrainian Academy of Sciences, pr. Nauki 47, Kiev, Ukraine}

 \begin{abstract}
The Dicke model extended to two bosons of different frequencies or
equivalent generalized Jahn-Teller lattice model are shown to
exhibit a spontaneous quantum phase transition between the
polaron-modified "quasi-normal" and squeezed "radiation" phase with
the transition point dependent on the frequencies. In a finite
lattice a mixed domain of coexistence of the quasi-normal and
modified radiation phase is created within the quasi-normal phase
domain. There occurs a field-directed oscillation-assisted tunneling
(hopping). The field is driven by simultaneous squeezing and
polaron-dressing of the collective boson level mode due to the
additional boson mode. In a finite lattice in the radiation domain
there occurs a sequence of local tunnelings (oscillations) between
two minima of a local potential weakly coupled to two assisting
oscillations. The "radiation" phase reveals itself as an almost
ideal instanton--anti-instanton gas phase. The correlations among
the energy levels mediated by the additional mode in the mixed
domain considerably reduce the level repulsions. As a consequence,
the Wigner level spacing probability distribution of the two-boson
Dicke model is non-universally reduced from the Wigner to the
semi-Poisson and asymptotically to the Poisson distribution of level
spacings.  The correlations cause a suppression of the coherence of
the radiation phase as finite-size effect. Possible applications of
the present theory are suggested.

\end{abstract}


 \maketitle

\section{Introduction}
\label{intro} The class of nonintegrable pseudospin-boson  lattice
models with one
\cite{Graham:1984:a,Graham:1984:b,Lew:1991,Cibils:1995,Dicke,Emary:2003:a,Emary:2003:b}
or two boson modes of different local symmetry
\cite{Haake,Tolkunov:2007,Pfeifer,Larson:2008} gained a long-term
interest in condensed matter physics and quantum optics. They
exhibit a rich variety of interesting statistical properties, e.g.,
quantum and thermodynamic phase transitions and quantum chaos in
energy spectra revealed in spectral statistics.

In the Dicke model \cite{Dicke} the quantum spontaneous phase
transition was found from the effectively unexcited "normal" phase
to the "super-radiant"  phase, a macroscopically excited and highly
collective state \cite{Emary:2003:a,Emary:2003:b}. A collective
spontaneous emission of coherent radiation is due to the cooperative
interaction of a large number of two-level atoms all excited to the
upper state (Dicke maser). The transition from quasi-integrability
to quantum chaos is given by a localization-delocalization
transition as a precursor of the phase transition
\cite{Emary:2003:b} in which the ground state wave function
bifurcates into a macroscopic superposition for any finite number of
atoms.
 Theoretical and experimental research of the superradiance or
superfluorescence of photons and spontaneous emission of phonons in
various variants of the Dicke-like collective system
(\cite{Skribanowitz:1973,Gross:1982,Andreev:1980,Florian:1984,
deVoe:1996}) still remains in a focus of interest
(\cite{Brandes:2000,Brandes:2005,Jarett:2007,Scully:2008}).

If a dephasing mechanism suppresses the strict coherence,
 processes based on collective spontaneous emission will reveal
 incoherent nature depending on the strength of the dephasing
and some peculiar effects result:
 Luminescence spectra of rare gas
solids, alkali halides (iodides) and some crystals based on
fullerenes comprise two contributions at low temperatures: a
broadband ascribed to selftrapped excitons and, at higher energies,
narrow-line (-band) optical transitions, fluorescence and stimulated
emission with sharply structured absorption and emission transitions
and a line-narrowing with inhomogeneous broadening due to free
excitons. Spontaneous emission reflected in fluorescence or
luminescence spectra was first observed as a free-exciton resonance
luminescence coexisting  with the broadband luminescence of the
selftrapped excitons excited  by electrons or synchrotron radiation
in rare gas solids and alkali halides (iodides) \cite{Kuusmann:1975}
and later by other authors \cite{Kmiecik:1987,Kishigami:1992}.
Transition from superfluorescence to amplified spontaneous emission
with lost coherence was studied on super-oxide ion in potassium
chloride in the intermediate regime as a function of temperature
\cite{Kishigami:1992}. Similar effects were observed also in
structures  based on Jahn-Teller (JT) active molecules, e.g.,
fullerenes  $C_{60}$ \cite{Ding:1997}. Laser-induced fluorescence
line narrowing
 experiment of $C_{60}$-hydroquinone crystals demonstrates
 inhomogeneous broadening of the zero-phonon lines. Similar observation
  has been done on Dicke superfluorescence
in KCl \cite{Malcuit:1987} and the broadening was ascribed to
unspecified internal strains. The peculiar high-energy narrow band
luminescence in the exciton and Jahn-Teller bearing structures
 has been proposed to be caused by the emission from the localized "exotic" excited states within the
 phonon-selftrapped spectrum by Wagner et al.  \cite{Kongeter:1990,Eiermann:1992}
(these states correspond to the tunneling  states proposed by
Slonczewski \cite{Slonczewski}).

Direct observation of the dynamic Jahn-Teller effect in $C_{60}$
\cite{Canton:2002} brought an evidence of important role of the
ionic motion in the electron-phonon interaction. The quantum
fluctuations led to formation of the tunneling states between two
energy wells of the distorted phase (selftrapped electrons) which
pertain to the undistorted geometry and recover free electron motion
found theoretically
\cite{Kongeter:1990,Eiermann:1992,Majernikova:2002}. This picture
corresponds to the concept of coexistence of selftrapped states and
tunneling to localized (exotic) states at higher energies.

 Physical interest in JT models is motivated by importance of some spatially anisotropic
complex structures (fullerenes, manganites, perovskites, etc) which
exhibit JT structural phase transition. The transition is due to
large local distortions related to the electron selftrapping. The JT
model is a prototype model for electron coupling to two degenerate
intramolecular vibrons of the same frequency (see (\ref{HJT}) below)
removing the degeneracy of electron levels ($\alpha$-term) and
phonon-mediated tunneling between the split levels ($\beta$-term).
Respective  vibron spectrum is complex with level crossings and
avoidings and needs exact diagonalization of  matrices of confined
level manifolds and proper statistical methods of evaluation.
  In crystals exhibiting high spatial anisotropy
 the rotation symmetry of JT molecules
 ($\alpha=\beta, \ \Omega_1=\Omega_2$, $\Omega_i$ being frequencies of the vibron modes)
 is generally broken. Therefore, it is
reasonable to investigate the JT model assuming different coupling
strengths for the onsite intralevel ($\alpha$) and interlevel
($\beta$) electron-phonon couplings and different vibron
frequencies. Such a model can also be considered as a generalization
of the exciton-phonon or dimer-phonon model, assuming the electron
tunneling to be phonon-assisted. It implies a competition of the
selftrapping  distortion mechanism (localization) and a  quantum
tunneling (delocalization) one.

We will show that the lattice JT Hamiltonian of molecules with
degenerate electron levels coupled with two non-equivalent boson
(vibron) modes is formally equivalent to the Dicke Hamiltonian
generalized by adding a boson mode coupled with the level spacing.
We shall investigate their excited spectra and related statistical
properties vs those of the one-boson Dicke model. An impact of the
additional boson mode which removes the double degeneracy of the JT
electron levels close to the quantum phase transition from normal to
supper-radiant phase reveals as a finite-size effect of level
correlations mediated by the additional boson. We will study
analytically an interplay of quantum fluctuations and symmetry
breaking  and numerically related characteristic signatures of a
quantum chaos - the level spacing probability distributions,
spectral entropies and spectral densities. We will demonstrate the
localization-delocalization transition of the wave functions within
the available lattice at the point of the phase transition between
the quasi-normal and radiation phase. As for applications, we shall
propose a zero temperature mechanism of the coherence dephasing and
the broadening of the spectra, including the broadening of the
zero-phonon lines. Further, we propose a similar impact on the
quantum localization-delocalization phase transition from the Mott
insulator to the superfluid state  of ultracold atoms in  optical
lattices.

Let us first shortly summarize the basic attributes  of the Dicke
model in order to motivate our use of the collective
Holstein-Primakoff bosonization method for the two-boson JT model.
The  exciton-boson (photon,
 phonon) model of a lattice chain of N atoms is represented by the Hamiltonian
\begin{equation}
 H= \sum_{k}\Omega_k a^{\dag}_k a_k + \omega_0\sum_{n=1}^N
\sigma_{z}^{(n)}
 +\lambda \frac{1}{\sqrt N}\sum_{n=1}^N\sum_{k}(a_k^{\dag}\exp(ikn)+a_k\exp(-ikn))
 \left( \sigma_+^{(n)} + \sigma_-^{(n)}
 \right),
\label{1}
\end{equation}
where $\sigma_i^{(n)}$ are spin variables; $n$ is a chain site, $N$
is the size of the chain, $k=\pm \pi l/N, \ l=0,1,\dots , N$ is the
wave vector of the phonon wave, $\Omega_k$ is the photon (phonon)
frequency of the $k$-th mode,  $\omega_0$ is the level spacing and
$\lambda$ is the level-phonon interaction strength. Phonon creation
and annihilation operators $a^+_k$, $a_k$ satisfy conventional
commutation rules $[a_k,a_k^+]=1$. The coefficient $1/\sqrt{N}$
arises from the dipole coupling strength \cite{Emary:2003:b}. This
factor allows also for the correct account of the transition to the
thermodynamic limit $N\rightarrow \infty$. The Dicke model is
obtained from the model (\ref{1}) when the boson is supposed
homogeneous (of long wave length) within a finite set of lattice
sites, $|k| \ll \pi$. Then, the collective spin variables $J_z=\sum
\limits_n \sigma_z^{(n)}$, $ J_+ +J_- = \dfrac{1}{\sqrt
N}\sum\limits_{n}\left( \sigma_+^{(n)} + \sigma_-^{(n)} \right)$
over all atoms in the chain can be introduced \cite{Dicke} and the
Dicke Hamiltonian yields
\begin{equation}
 H= \Omega  a^{\dag} a + \omega_0 J_{z}
 +\lambda  \frac{1}{\sqrt N} (a^{\dag}+a)( J_+ + J_- )\,.
 \label{2}
\end{equation}

 The collective (N-dimensional)
spin variables $J_z$, $J_+$, $J_-$ satisfy the same commutation
relations as the individual spins:
\begin{equation}
 [J_z, J_{\pm}]=\pm J_{\pm}\,; \quad [J_+,J_-]=2 J_z
\label{spin:alg}
\end{equation}
with $J_x=\dfrac{1}{2}(J_+ +J_-), J_y=-\dfrac{i}{2}(J_+-J_-)$. It is
handy to represent the realization of the spin operators of higher
dimensions by the Dicke states. Each Dicke state is labeled as
$|j,m\rangle$ with the discrete index $m$ ranging (at fixed $j$)
from  $-j,-j+1,\dots$ to $j$. The spin operators in the Dicke space
are defined as follows:
\begin{equation}
 J_z |j,m\rangle = m |j,m\rangle\,; \quad J_{\pm}|j,m\rangle= \sqrt{j(j+1)-m(m\pm 1)}
 |j,m\pm 1\rangle\,; \quad J^2 |j,m\rangle=j(j+1) |j,m\rangle\,.
\label{Dicke:alg}
\end{equation}
The subspaces of the Dicke states labeled by $j$ are independent
and, hence, one can consider them separately keeping $j$ fixed.
 For the chain containing $N$ atoms possible values of $j$ range as $0,\, 1,\dots , N/2$ for
 $N$ even and $1/2,\, 3/2, \dots , N/2$ for  $N$ odd; for given $j$ the operators $J_i$ are matrices
 with dimensions $2j+1$ (representations of the $SU(2)$ group).
The parity operator
\begin{equation}
\Pi = \exp \{i\pi (a^{\dag}a + J_z+j)\},
 \label{parity}
\end{equation}
  commutes with Hamiltonian (\ref{2}), i.e.
 the number of excitation quanta  $ \langle a^{\dag}a \rangle + J_z+j$ is conserved within a
 phase.  Namely, the space
of states is separated onto two subspaces with the number of
excitation quanta which is either even or odd
\cite{Emary:2003:a,Emary:2003:b}. Then the split subspaces can be
considered separately. This conservation is broken at the quantum
phase transition.

Two-level pseudospin-vibron (photon) JT model Hamiltonian
investigated in
\cite{Majernikova:2003,Majernikova:2006:b,Majernikova:2006:a} has
the form of the $2\times 2$ matrix which in the pseudospin notation
is written as

\begin{equation}
 H= \Omega (a_{1}^{\dag}a_{1} +{a_2}^{\dag}a_{2}+1 )I + \alpha
(a_{1}^{\dag}+a_{1})\sigma_{z}
 -\beta ({a_2}^{\dag}+a_2)\sigma_{x}
\label{HJT}
\end{equation}

Hamiltonian (\ref{HJT}) describes the electron-boson interaction in
a two-level JT molecule with two boson modes $a_i$ of equal
frequencies $\Omega$ ($I$ in the first term is the unity matrix). In
(\ref{HJT}), the $\alpha$-coupled mode lifts the double degeneracy
of the local JT level and $\beta$-coupled mode mediates the electron
tunneling between the split levels \cite{Long:1958}. Different
electron-boson coupling strengths $\alpha$ and $\beta$ present a
generalization with reference to the conventional (rotational
symmetric) JT model where  $\alpha=\beta$ \cite{Majernikova:2006:a}.

In a similar way as for the exciton model (\ref{1}) a natural
generalization of the two-level pseudospin-vibron (photon) JT
Hamiltonian can be done by extending the dimensionality of the
(pseudo)spin space. Namely, the generalized model considered within
the present work is characterized by the following Hamiltonian:

\begin{equation}
H =
 \Omega_1 \left(\frac{1}{2} + a_{1}^{\dag}a_{1} \right)I +  \Omega_2 \left(\frac{1}{2} + a_{2}^{\dag}a_{2} \right)I
 + \frac{\alpha}{\sqrt{2j}}
(a_{1}^{\dag}+a_{1})\cdot 2J_{z}
 +\frac{\beta}{\sqrt{2j}} (a_2^{\dag}+a_2)\left( J_+ + J_- \right)\,.
\label{H:gener}
\end{equation}

The operators $J$ are  defined in representation of the $SU(2)$
group (\ref{spin:alg}) in a $2j+1$-dimensional space spanned by
Dicke vectors $|j,m\rangle$ as described above
(Eq.(\ref{Dicke:alg})). We assume a general case of the rotation
symmetry breaking with different boson frequencies $\Omega_1$ and
$\Omega_2$.

 In the molecular JT model the bosons represent the
intramolecular vibrons. The rotational symmetry is broken by taking
different frequencies and interaction parameters $\alpha\neq \beta$.
The generalization acquires  a twofold interpretation: (i) The JT
lattice model, $j$ lattice sites represent a finite number of
identical JT molecules, i.e. two-level molecules coupled to two
intramolecular vibrons (of the same intramolecular structure). Then
the electron collective coordinate can be introduced representing a
collective spin of $N$ two-level molecules (ensemble of qubits)
coupled with two local vibron modes identical for each on-site
molecule.
 (ii) The excited on-site molecule with two intra-molecular vibrons can be
considered by taking a set of many excited ($j$) levels of the
molecule. (\ref{H:gener}) can be then interpreted as usual
many-level system (a single atom with a spin $\geq 1/2$) coupled to
two boson modes.

The above concept of JT model can be applied, e. g., for the
formulation of the JT polaron model in nanostructures where due to
their finite size the quantum effects are rather detectable for
observation.

 Additionally the photonic JT model
\cite{Pfeifer,Larson:2008} with collective boson modes homogeneous
within the chain of atoms is within the scope of (\ref{H:gener}) as
well.

 In the Dicke model, two bosons represent optical fields of the wave length
larger than the size of the lattice and so they enable to introduce
the collective variables for the electron levels. Mathematically,
both the generalized Dicke and JT models are then equivalent and
represented by equation (7).

Thus an impact of the coupling of the level spacing with the
additional boson  $a_1$ compared to the conventional one-boson Dicke
model (\ref{2}) is to be studied.

Likewise in the exciton model there exists the same operator of
parity $ \Pi = \exp \{i\pi (a_2^{\dag}a_2 + J_z+j)\}$, which
commutes with Hamiltonian (\ref{H:gener}).
 For the JT molecule  with
only two electron levels  ($j=1/2$) and equal boson frequencies this fact reflects itself in the conserved
parity $p=\pm 1$ \cite{Long:1958,Majernikova:2003} as an
additional good quantum number.  For parities $+1$ and $-1$ the
spectra of that model are identical. As we shall see, for the
present generalization this is the case of even number of molecules,
that is for the main Dicke numbers $j=1/2,3/2,\dots$.

Finally, a diagonal term of the form $\Delta J_z$ can be added to
the (\ref{H:gener}) where the parameter $\Delta$ accounts for the
intrinsic level separation. This term accounts for a magnetic field
\cite{Larson:2008}. In the JT model, this term is known to affect
the shape of the adiabatic potential: it lifts the conical
intersection of upper and lower adiabatic surfaces.

Number and relative values of the pseudospin-boson couplings
involved  are of essential importance for specification of
statistical properties of the pseudospin-boson models: while the
excited spectra of various pseudospin-one-boson models were found to
exhibit Wigner chaos when approaching the semiclassical limit
\cite{Graham:1984:a,Graham:1984:b,Lew:1991,Cibils:1995,Emary:2003:a,Emary:2003:b},
the excited spectra of the molecular  JT  model
($\Omega_1=\Omega_2$) exhibit nonuniversal chaotic behaviour between
the Poisson and semi-Poisson limit
\cite{Majernikova:2006:b,Majernikova:2006:a, Majernikova:2008}. The
antisymmetric against reflection level splitting mode $a_1$ of the
JT model increases by one the number of local degrees of freedom
when compared to the exciton model. Simultaneously, this phonon
(photon) mode mediates additional level correlations and quantum
fluctuations and thus it weakens the level repulsions. The
interaction effects are modified by the frequency difference  and
are expected to imply an impact on the statistical properties of the
lattice models such as quantum phase transition to the super-radiant
phase and proper statistical characteristics of the quantum chaos.

In Section II we investigate numerically and analytically the ground
state and related phase transition of the extended Dicke model or,
equivalently, of the lattice JT model (\ref{H:gener}) with two
bosons of different frequencies. For analytical calculations we use
the Holstein-Primakoff bosonization Ansatz for the pseudospin
variables. We analyze the macroscopic phases for three involved
oscillators related to the competing correlations (self-trapping and
squeezing) due to the additional mode. In Section III we present
analytical investigations of the interplay of quantum fluctuations
and the symmetry breaking  as a useful insight into the scenario
leading to the  macroscopic phase transition in the system of three
oscillators.  For a finite lattice,  within the quasi-normal domain
a medium quantum domain is shown to be created in which both the
"normal" and "radiation" phase coexist. In Section IV we present
numerical results on wave functions and statistical properties of
excited states, namely level spacing probability distributions for
various number of lattice sites, values of frequencies and coupling
parameters and compare them with those of the one-boson Dicke model.
Further, the distribution of occupation of excited states over the
lattice sites is illustrated by the state entropies as functions of
 quantum numbers of even states. The spectral densities as functions of
 energy are calculated as illustration of the effect of the
 interaction with the boson modes on the spectral density on the Dicke spectral space.
 The wave functions and the statistical characteristics reflect the
 medium quantum domain as being mixed from the localized
 and delocalized contributions of the normal and radiation phase,
 respectively.

For presentation of some numerical results we use the resonance case
$\Omega_1=\Omega_2$ except for the cases where the difference of the
frequencies makes qualitative changes.
 Throughout all the paper we shall use the term
 "radiation" phase  for our model with suppressed coherence
 instead of the term  "super-radiant" standardly used for the  strictly coherent one-boson Dicke model.

\section{Ground state and phase transition}

Use of the method of collective pseudospin variables for two-level
electron-boson JT and Dicke lattice models is appropriate under the
conditions presented in Introduction.  The collective coordinates
enable solving these nonintegrable models efficiently, though
approximately. In what follows, we apply them for the investigation
of a lattice of JT molecules or the two-boson Dicke model.

 Hamiltonian of the system can be  written as a generalized
version of (\ref{1}) ($k=0$) as follows
\begin{equation}
H=
\Omega_1 \left(\frac{1}{2} + a_{1}^{\dag}a_{1} \right)I +
\Omega_2 \left(\frac{1}{2} + a_{2}^{\dag}a_{2} \right)I +
\sum\limits_{i=1}^N
\frac{1}{\sqrt {2j}} \Big[\alpha (a_1^{\dag}+a_1)2 s_z^{(i)}+\beta
(a_2^{\dag}+a_2)(s_{+}^{(i)}+s_{-}^{(i)})\Big]\,,\label{h}
\end{equation}
where $j$ assumes values in different subspaces $j=0,1,2,\dots N/2$
for $N$ even and $j=1/2,\,3/2,\dots N/2$ for $N$ odd. In what
follows  we will always take into consideration only the subspace
with largest $j=N/2$ for each number of atoms in a  chain. The
frequencies of both modes are $\Omega_1$, $\Omega_2$ and
corresponding coupling constants are $\alpha$, $\beta$. Introducing
 the  collective coordinates $J_i$ defined by
 (\ref{spin:alg}) into equation (\ref{h}) in the same fashion as for (\ref{1}) we arrive at
Hamiltonian (\ref{H:gener}) representing either the generalized
 Dicke or JT lattice model. For $j=1/2$, the local (molecular) JT
model (with broken rotational symmetry if $\alpha\neq\beta$,
$\Omega_1\neq\Omega_2$) is recovered. For completeness, the standard
Dicke-like model is the one-boson version of Hamiltonian
(\ref{H:gener}) without assistance of the bosons $1$.

For analytical treatment it is convenient to bosonize the spin
variables in the two-boson-spin system (\ref{H:gener}) using the
Holstein-Primakoff Ansatz \cite{Primakoff}

 \begin{equation}
  J_z= b^{\dag}b-j,\ J_+=
b^{\dag}\sqrt{2j-b^{\dag}b}, \ J_-= \sqrt{2j-b^{\dag}b}\,b
\label{hp}
 \end{equation}
  for the collective pseudospin operators. Here, $b$ are fictitious
  boson operators, $[b,b^\dag]=1$. This representation of the spin
algebra preserves exactly the commutation relations (\ref{spin:alg})
and makes it possible to convert the system to two or three coupled quantum
oscillators as will be shown below.

Linear terms in the Hamiltonian (\ref{H:gener}) can be excluded by
the use of the coherent state representation with boson
displacements chosen variationally. Let us displace the operators
involved as follows
\begin{equation}
 a_1^{\dag}=
c^{\dag}_1+\sqrt{\alpha_1}, \quad a_2^{\dag}=  c^{\dag}_2+
\sqrt{\alpha_2},   \quad
b^{\dag}= d^{\dag}-\sqrt{\delta}\,.\label{shift}\end{equation}

Applying the Holstein-Primakoff Ansatz (\ref{hp}), setting
(\ref{shift}) into  (\ref{H:gener}) and  expanding the square root
expressions in (\ref{hp}) to first order terms in $d^{\dag}d$ yields
the form
\begin{align}
&H= \Omega_1\left( \frac{1}{2} +c_1^{\dag}c_1 \right) + \Omega_2\left( \frac{1}{2} +c_2^{\dag}c_2 \right)
+ \Omega_1\alpha_1+\Omega_2\alpha_2 +\Omega_1
\sqrt{\alpha_1}\left(c_1^{\dag}+c_1\right)+\Omega_2
\sqrt{\alpha_2}\left(
c_2^{\dag}+ c_2\right)\nonumber\\
& +\frac{2\alpha}{\sqrt
{2j}}\left[(c_1^{\dag}+c_1)d^{\dag}d +2\sqrt{\alpha_1}d^{\dag}d
-\sqrt{\delta}(c_1^{\dag}+c_1)(d^{\dag}+d)-2\sqrt{\alpha_1\delta}(d^{\dag}+d)+
(\delta-j)(c_1^{\dag}+c_1) \right.\nonumber\\
& +2\sqrt{\alpha_1}(\delta-j)\Big]+ \frac{
\beta}{\sqrt{2j}}( c_2^{\dag}+ c_2+2\sqrt{ \alpha_2})
\cdot k \left \{-2\sqrt{\delta} +
(1-\delta/k^2)(d^{\dag}+d)+\frac{\sqrt{\delta}}{k^2}d^{\dag}d\right.  \nonumber\\
&\left. +\frac{1}{2k^2}\left[-(d^{\dag
2}d+d^{\dag}d^2)+\sqrt{\delta}((d^{\dag}+d)^2-1)
 \right ]\right\},\quad k\equiv\sqrt{2j-\delta}\,. \ \
\label{H}
\end{align}

From the condition of elimination of the terms linear in boson
operators from (\ref{H}) three identities for the parameters
$\alpha_i$ and $\delta$ follow:
 \begin{align}
&\Omega_1\sqrt{\alpha_1}= -\frac{ 2\alpha}{\sqrt{2 j} }\,(\delta-j)\,, \nonumber \\
&\Omega_2 \sqrt{\alpha_2}= \frac{\beta\sqrt{\delta}}{\sqrt{2j}}\,
\frac{4j-2\delta+1/2}{\sqrt{2j-\delta }}\,, \label {displ} \\
&\sqrt{\delta}(\delta-j)\left
(4\Omega_2\,\alpha^2-2\Omega_1\,\beta^2\,\frac{4j-2\delta+1/2}{2j-\delta}\right )=0 \,.
\nonumber
\end{align}

It we consider a finite lattice ($j$ finite), this set of equations
implies three solutions provided $ \Omega_2\,\alpha^2\neq \Omega_1\, \beta^2(1+1/8j)$
as follows:
\begin{align}  & 1. \  \sqrt{\alpha_1}=\frac{\alpha}{\Omega_1}\sqrt {2j}, \ \alpha_2=0 , \, \delta=0\
, \label{sol1}\\
 & 2.  \   \alpha_1= 0, \  \sqrt
{\alpha_2}=\frac{\beta}{\Omega_2}\sqrt{2 j}\left(1+\frac{1}{4j}\right) , \  \delta=j  ,  \label{sol2} \\
& 3.  \ \sqrt{\alpha_1}= -\frac{\alpha}{\Omega_1}\sqrt{2j}
(1-2\bar\mu) , \ \sqrt{\alpha_2}=\frac{\alpha^2}{\Omega_1}
\frac{2\sqrt{2j}}{\beta} \sqrt{\bar{\mu}(1-\bar{\mu})}, \
\delta=2j(1-\bar\mu )\,, \label{sol3}
\end{align}
 where in the last set  $\bar\mu = \dfrac{\beta^2\, \Omega_1}{8j\Big(\alpha^2\,\Omega_2 - \beta^2\,\Omega_1\Big)}
\equiv \dfrac{\bar{\beta}^2}{8j\left(\alpha^2-{\bar{\beta}}^2\right)}
 <1$, and
$\alpha > \bar{\beta} \equiv \sqrt{\Omega_1/\Omega_2}\,\beta$.

As will be seen in the next subsections, first two solutions in the
thermodynamic limit $j\to\infty$ account for the common ``normal''
and ``super-radiant'' phases of a system, the transition between
them at $\alpha\sqrt{\Omega_2}=\beta\sqrt{\Omega_1}$ describing the
quantum phase transition (QPT) similar to what was reported for the
Dicke models of superradiance
\cite{Emary:2003:a,Emary:2003:b,Brandes:2005,Tolkunov:2007}. The
third solution  in the thermodynamic limit becomes trivially
degenerated and valid only on the line of the said QPT. Nontrivial
consequences for this solution arise for finite $j$ only: From Eqs.
(\ref{sol1}) and (\ref{sol2}) a transition to the rotational
symmetry of the two-dimensional vibron oscillator is evident at $
\alpha_2=\alpha_1$ at the critical point
$\alpha=\beta\sqrt{\dfrac{\Omega_1}{\Omega_2}}(1+1/4j)$. In the
limit of large $j\rightarrow\infty$ this is the point of the phase
transition described below. This limit of higher symmetry exhibits
qualitatively different properties when compared with the case
$\alpha\neq\beta\sqrt{\Omega_1/\Omega_2}(1+1/4j)$ and $j$ finite as
we shall see in Section IV. Let us note, that the factors like $\sim
1/j$ are the origin of specific quantum effects such as the
squeezing of the frequency $\omega_{2}= \dfrac{4{
\beta}^2}{\Omega_2}\cosh (4r)$, $\cosh 4r= 1+1/4j $, of the
oscillator (polaron) $2$ (subsection B). The effect of squeezing
gets implications for the level spacing probability distributions
especially in the case 3 when all three oscillators are coupled
(Subsection C).

\subsection{Case 1: Polaron self-trapping in the normal phase.}

 The solution  (\ref{sol1}) represents  the normal phase with average zero number of macroscopically excited
electron level bosons $\delta=0$.
  Hamiltonian (\ref{H}) related to this solution reads
\begin{multline}
H_1= \Omega_1\left( \frac{1}{2} +c_1^{\dag}c_1 \right) +
\Omega_2\left( \frac{1}{2} +c_2^{\dag}c_2 \right)
+\frac{4\alpha^2}{\Omega_1}(d^{\dag}d-j/2)+ \beta (
{c_2}^{\dag}+  c_2)(d^{\dag}+d) \\
+\frac{2\alpha}{\sqrt{2j}}(c_1^{\dag}+c_1)d^{\dag}d -\frac{
\beta}{4j }( c_2^{\dag}+c_2)(d^{\dag 2} d +d^{\dag } d^2)\,.
\label{H1}
\end{multline}

Here,  the level oscillator  acquires the frequency of a polaron-
dressed oscillator $\omega_1=4\alpha^2/\Omega_1$ which results from
the self-trapping by the additional mode in contrast with respective
factor of the Dicke model  $\omega_0  $.
 Explicit interaction of the
oscillator modes with the level bosons contribute only to excited
states  $\propto 1/\sqrt j $ in (\ref{H1}). Thus, in the limit
$j\rightarrow \infty$ Hamiltonian (\ref{H1}) recovers the normal
phase Hamiltonian of polaron level oscillators. For large $j$ only
linear terms persist (first line in (\ref{H1}))  and it can be easily diagonalized
by a rotation in the plane of coupled operators $c_2$ and $d$
\cite{Emary:2003:b} to yield the form
\begin{equation}
 H_{1d}= \Omega_1 c_1^{\dag}c_1+ \epsilon_{1,
+}C_2^{\dag}C_2 +  \epsilon_{1,
-}C_3^{\dag}C_3-\frac{2\alpha^2}{\Omega_1}j\, \label{E1}
\end{equation}
with new effective oscillators $C_2$, $C_3$.
The excitation energies of the system $\epsilon_{1,\pm}$ are

\begin{equation}
\epsilon_{1, \pm}^2=
\frac{1}{2}\Big(\Omega_2^2 +\omega_1^2\pm
\left[(\Omega_2^2-\omega_1^2)^2+ 64\alpha^2 \beta^2\frac{\Omega_2}{\Omega_1} \right]^{1/2}\Big)\,.\label{eps1}
\end{equation}
 From (\ref{eps1}), the solution for $\epsilon_1 $ exists
provided $\omega_1> 4\beta^2/\Omega_2 \equiv \omega_2$, or if
$\alpha\sqrt{\Omega_2} > \beta\sqrt{\Omega_1}$. This
 phase is identified as the normal phase of the Dicke-like
model without macroscopic excitations, Fig. \ref{boson-fig1}. The effect of
different initial frequencies $\Omega_1\neq \Omega_2$ is just the
shift of the phase transition point supporting the phase with the
smaller of $\Omega_i$.

From (\ref{H1}),  critical Hamiltonian $H_1^{crit}= H_1
(\beta_c=\alpha\sqrt{\Omega_2/\Omega_1})$ at the point of the phase
transition yields
\begin{equation}
H_1^{crit}=\sqrt{\Omega_2^2+
\left(\frac{4\alpha^2}{\Omega_1}\right)^2} \left(C^{\dag}_2
C_2+1/2\right) + \Omega_1 (c_1^{\dag}c_1+1/2) -\frac{2\alpha^2
j}{\Omega_1}\,. \label{H1crit}
\end{equation}

  The coupled undisplaced oscillators $a_2$ and $b$ form an
effective single oscillator of the frequency
$\sqrt{\Omega_2^2+\left (4\alpha^2/\Omega_1\right )^2}$. The mixing
results from the coupling due to the assisting oscillator $a_2$
between the levels split by the oscillator $a_1$. The additional
coherent oscillator $c_1$ remains decoupled.

\subsection{Case 2: Squeezing in the radiation phase.}

The second solution (\ref{sol2}) identified further with the
superradiant phase, can be treated on the same footing. For this
solution the level bosons (\ref{hp}) are displaced by $\delta=j$
that is they acquire a macroscopic number of excited quanta.
 Respective  Hamiltonian (\ref{H}) yields
\begin{align}
& H_2=\Omega_1\left( \frac{1}{2} +c_1^{\dag}c_1 \right) +
\Omega_2\left( \frac{1}{2} +c_2^{\dag}c_2 \right)
+\frac{2\beta^2}{\Omega_2}\left(1+\frac{1}{4j}\right) \left(d^{\dag}d-j-\frac{1}{4}\right)
-\sqrt{2}\alpha (c_1^{\dag}+c_1)(d^{\dag}+d)\nonumber\\
& +
\frac{\beta^2}{\Omega_2}\left(1+\frac{1}{4j}\right)(d^{\dag}+d)^2 +
\frac{2\alpha}{\sqrt{2j}}(c_1^{\dag}+c_1)d^{\dag}d
+\frac{\beta}{\sqrt{2j}}(c_2^{\dag}+c_2)d^{\dag}d
+\frac{\beta}{2\sqrt {2 j}}(c_2^{\dag}+c_2)\nonumber\\
& \times \left[(d^{\dag}+d)^2 -\frac{1}{\sqrt j}(d^{\dag 2}d
+d^{\dag}d^2 )\right]-\frac{\beta^2}{\Omega_2\sqrt{j}}\left(1+\frac{1}{4j}\right)\left(d^{\dag
2}d +d^{\dag}d^2 \right)\, . \label{H2}
\end{align}

Linear part of (\ref{H2}) relevant in approaching the thermodynamic
limit $j\to\infty$ represents the level polaron coupled with the
oscillator $1$. Quantum fluctuations squeeze its frequency by the
squeezing parameter ${\rm cosh}\ 4r= (1+1/4j)$. Transformation of
this part of (\ref{H2}) by the unitary operator of squeezing $S=
\exp [r(d^{\dag 2}-d^2)]$ using the identities
\begin{align}
&\tilde{d^{\dag}}\tilde d\equiv  S  d^{\dag}d S^{-1}= d^{\dag} d\cosh
4r + \sinh^2 2r + (d^2+d^{\dag 2})\frac{1}{2}\sinh 4r, \nonumber \\
&(\tilde{d^{\dag}}+\tilde d) \equiv
S(d^{\dag}+d)S^{-1}=(d^{\dag}+d)e^{-2r} \label{S}
 \end{align}
 yields the
squeezed polaronic oscillator  with renormalized frequency $
\omega_2=4\beta^2/\Omega_2\cosh (4r)$ and interaction $\kappa=\sqrt
2\alpha e^{2r}$,
\begin{equation}
\tilde{H_2}=
\Omega_1\left( \frac{1}{2} +c_1^{\dag}c_1 \right) +
\Omega_2\left( \frac{1}{2} +c_2^{\dag}c_2 \right)
+ \omega_2
 \tilde d^{\dag} \tilde d-\kappa (c_1^{\dag}+c_1)( \tilde d^{\dag}+ \tilde
 d)-\frac{2\beta^2
}{\Omega_2} j-\omega_2 \sinh^2 2 r\,. \label{tilde}
\end{equation}
 Diagonalization  of (\ref{tilde})  implies three independent
 oscillators $C_1, C_3$ and $c_2$, last one remaining free,
 \begin{equation}
 H_{2d}= \epsilon_{2,
+}C_3^{\dag}C_3 +  \epsilon_{2, -}C_1^{\dag}C_1+\Omega_2
c_2^{\dag}c_2 -\frac{2\beta^2}{\Omega_2}j \,, \label{E2}
\end{equation}
where
\begin{equation}
\epsilon_{2, \pm}^2= \frac{1}{2}\left(\Omega_1^2+ \omega_2^2\pm
\Big[(\Omega_1^2-\omega_2^2)^2+ 64 \alpha^2 \beta^2 \e^{4r} \frac{\Omega_1}{\Omega_2}
\Big]^{1/2}\right)\,. \label{eps2}
\end{equation}
 Energy of the
squeezed ground state is $H_{2G}= -2\beta^2/\Omega_2\cdot
j-\omega_2\sinh^2 2r.$ From (\ref{eps2}), one obtains the condition
for stability of the radiation phase
\begin{equation}
\frac{\beta^2}{\Omega_2} > \frac{ \alpha^2}{\Omega_1} e^{4r}.
\label{beta>alpha}
\end{equation}
Let us remind that for $j\rightarrow\infty $ the squeezing parameter
$r\rightarrow 0 \  (\sinh 4r\sim 1/\sqrt{2j})$.

Close to the phase transition point
$\beta_c=\alpha\sqrt{\Omega_2/\Omega_1}$ where $\epsilon _1^{(-)}=
0$, we can find the energy from the side of the normal phase
$\alpha\rightarrow \beta_{c-}$
\begin{equation}
\epsilon^{(-)}\rightarrow
\frac{8\,\alpha^{3/2}\Omega_2^{3/4}}{\sqrt{2}\,\Big(\Omega_1^2\Omega_2^2+16\alpha^4\Big)^{1/2}}
\left(\alpha\sqrt{\Omega_2}-\beta\sqrt{\Omega_1}\right)^{1/2}\,.
\label{epscrit1}
\end{equation}
 From the side of the
radiation phase, at $\beta_{c+}\leftarrow \alpha_+ $,   the energy
reads
\begin{equation}
\epsilon^{(+)}\rightarrow
\frac{8\,\beta^{3/2}\Omega_1^{3/4}\e^{2r}}{\sqrt{2}\,\Big(\Omega_1^2\Omega_2^2+16\beta^4\Big)^{1/2}}
\left(\beta\sqrt{\Omega_1}-\alpha\sqrt{\Omega_2}\right)^{1/2}.
\label{epscrit2}
\end{equation}
 In the limit $j\rightarrow \infty$ the squeezing parameter $r$ vanishes, and
  the energy dependence becomes symmetric with simultaneous interchange $\alpha\leftrightarrow\beta$,
$\Omega_1\leftrightarrow\Omega_2$.  Note that in the resonance case
$\Omega_1=\Omega_2$ the  cusps in  (\ref{epscrit1}),
(\ref{epscrit2}) are symmetric around the critical point
$\alpha=\beta$ (Fig. \ref{boson-fig1}). However, for finite $j$ this
symmetry is broken because of the factor  $\exp(2r)$
 in the numerator of (\ref{epscrit2}): the radiation phase (\ref{epscrit2}) is suppressed by the
squeezing when compared to the normal phase. The non-symmetry of the
branches below and above the
 critical point is also evident from numerical results for the excitation energy
 from the exact Hamiltonian (\ref{H:gener}) presented in Fig. \ref{boson-fig1}. Let us note,
that the critical exponent $1/2$ near  the critical point in
(\ref{epscrit1}) and (\ref{epscrit2}) is  characteristic for the
mean field second order phase transition. The branches
(\ref{epscrit1}) and (\ref{epscrit2}) represent a generalized
version of the branches by Emary and Brandes \cite{Emary:2003:b}:
The analogy becomes obvious if we set
 $\lambda_c=\frac{1}{2}\sqrt{\omega\omega_0}\equiv\sqrt{\dfrac{\Omega_1}{\Omega_2}}\beta_c$, with $\omega\equiv\Omega_1=\Omega_2$, $\omega_0\equiv
4\alpha^2/\Omega_1$.

 In the limit $j\rightarrow\infty$,
similar critical behavior as (\ref{H1crit}) can be found for the
energy of the phase $2$,  $H_2^{crit}= H_2 (\sqrt{\dfrac{\Omega_1}{\Omega_2}}\beta_c=\alpha)$.
From (\ref{E2}), one obtains
\begin{equation}
H_2^{crit}=\sqrt{\Omega_1^2+  \left(\frac{4\beta^2}{\Omega_2}\right)^2} (C^{\dag}_1
C_1+1/2) + \Omega_2 (c_2^{\dag}c_2+1/2) -\frac{2\alpha^2
}{\Omega_1}j, \label{H2crit}
\end{equation}
where now the displaced oscillator $c_2$ remains decoupled and the
undisplaced oscillator $c_1$ mediates a mixing with the split level
polaronic oscillators of the frequency $\omega_2=4\beta^2/\Omega_2$.

 The analysis of both phases shows that within linear
approximation the oscillator $c_1$ in the radiation phase plays
qualitatively the same role as it does the oscillator $c_2$ with
simultaneous interchange of the polaron frequencies $\omega_{1}
\leftrightarrow\omega_{2}$ (coupling constants $\alpha
\leftrightarrow \beta\sqrt{\frac{\Omega_1}{\Omega_2}}$. Hence, in
linear approximation two oscillators mix to two-dimensional
effective oscillator while the remaining one is decoupled. For the
resonant case $\Omega_1=\Omega_2$ and $\alpha=\beta$
($\omega_1=\omega_{2}$) a spontaneous transition occurs (in the
plane of vibron coordinates $(Q_1,Q_2)$) to the rotation symmetric
two-dimensional oscillator of a qualitatively different behavior
which is out of the scope of this paper. This case implies an abrupt
change of the statistical characteristics of the quantum excited
levels (section IV) related to the higher symmetry of the problem
\cite{Majernikova:2006:b}.

\subsection{Case 3. Intermediate
domain of mixed quasi-normal and "radiation" phase }

 The solution (\ref{sol3}), valid for finite $j$, implies for  Hamiltonian (\ref{H}) following
 form
\begin{align}
& H_3 =
\Omega_1\left( \frac{1}{2} +c_1^{\dag}c_1 \right) +
\Omega_2\left( \frac{1}{2} +c_2^{\dag}c_2 \right)+
\frac{4\alpha^2\bar{\mu}}{\Omega_1}d^{\dag}d - \nonumber \\
& 2\alpha\sqrt{1-\bar{\mu}}(c_1^{\dag}+c_1)(d^{\dag}+d)
+\beta\frac{2\bar{\mu}-1}{\sqrt{\bar{\mu}}}
(c_2^{\dag}+c_2)(d^{\dag}+d) + \nonumber \\
& \frac{2\alpha^2}{\Omega_1} (1-\bar{\mu})
(d^{\dag}+d)^2
+ \frac{2\alpha}{\sqrt{2j}}(c_1^{\dag}+c_1)d^{\dag}d
+\frac{\beta}{\sqrt{2j}}
\frac{\sqrt{1-\bar{\mu}}}{\sqrt{\bar{\mu}}}(c_2^{\dag}+c_2)d^{\dag}d \nonumber  \\
& +\frac{\beta}{2\sqrt {2j}}
\frac{\sqrt{1-\bar{\mu}}}{\sqrt{\bar{\mu}}}
 (c_2^{\dag}+c_2)(d^{\dag}+d)^2 \nonumber\\
& -  \frac{2\alpha^2}{\Omega_1\sqrt
{2j}}(1-\bar{\mu})^{1/2}(d^{\dag
 2}d+d^{\dag}d^2)-\frac{\beta}{4j\sqrt{\bar{\mu}}}(c_2^{\dag}+c_2)(d^{\dag
2}d+d^{\dag}d^2) - \nonumber \\
&\frac{2\alpha^2}{\Omega_1}j(1-2\bar\mu)^2+\frac{8\alpha^4 \Omega_2 \,
j}{\beta^2\, \Omega_1^2}\bar{\mu}(1-\bar{\mu})
-\frac{2\alpha^2(1-\bar\mu)}{\Omega_1} -
\frac{16\alpha^2j\bar{\mu}(1-\bar{\mu})}{\Omega_1}\,,  \label{H3}
\end{align}
where $\bar{\mu}=\dfrac{\beta^2\,\Omega_1}{8j\Big(\alpha^2\,
\Omega_2 -\beta^2\, \Omega_1 \Big)} <1$, $\alpha > \beta
\sqrt{\dfrac{\Omega_1}{\Omega_2}}\,$.

From (\ref{H3}) one obtains for the quasi-classical oscillators
$\langle c_1^{\dag}+c_1\rangle\equiv  q_1\sqrt {2\Omega_1} $,
$\langle c_2^{\dag}+c_2\rangle \equiv q_2\sqrt {2\Omega_2} $ and for
the level polaron $\langle d^{\dag}+d\rangle= Q
\sqrt{8\alpha^2\bar{\mu}/\Omega_1 } $  dynamic equations
\begin{align}
&\ddot{q_1}= -\Omega_1^2q_1+4\alpha
\sqrt{1-\bar{\mu}}\sqrt{\omega\Omega_1}Q-\frac{\alpha}{\sqrt
j}\omega\sqrt{\Omega_1}Q^2,\\
&\ddot{q_2}= -\Omega_2^2q_2- 2\beta
\frac{2\bar\mu-1}{\sqrt{\mu}}\sqrt{\Omega_1\omega}Q-
\frac{\beta\omega\sqrt{\Omega_2}\sqrt{1-\bar\mu}}{2\sqrt{j\bar\mu}}Q^2,\\
&\ddot{Q}= -\omega^2 Q+4\alpha
\sqrt{1-\bar\mu}\sqrt{\omega\Omega_1}q_1-2\beta\frac{2\bar{\mu}-1}{\sqrt{\mu}}\sqrt{\omega\Omega_2}q_2
-\frac{4\alpha^2}{\Omega_1}\frac{1-\bar{\mu}^2}{\bar{\mu}}\omega Q\nonumber\\
&-\beta\sqrt{\frac{1-\bar\mu}{j\bar{\mu}}}\omega\sqrt{\Omega_2}q_2Q-\frac{\beta}{\sqrt
j}\frac{1-\bar{\mu}^2}{{\bar{\mu}}^2}\omega\sqrt{\Omega_2}q_2Q\nonumber\\
&+\frac{6\alpha^2\omega}{\Omega_1\sqrt {2j}}(1-\bar\mu)^{1/2}Q+
\frac{3\beta\omega\sqrt{2\Omega_2}}{4j\sqrt{\bar\mu}}q_2 Q,
.\label{6}
\end{align}

Linear approach used for the normal and radiation phase far from the
critical region can not be applied for the mixed critical region;
Instead, nonlinear dynamic equations are to be solved.
 The dynamics of the level oscillator $Q$ in (\ref{6}) is
  influenced by the squeezing parameter $\bar{\mu}$:
  If $j$ is sufficiently large,
  $\bar{\mu}$ is small
  ($\alpha^2>\beta^2\Omega_1/\Omega_2$)
 and  $\omega=4\alpha^2\bar{\mu}/\Omega_1 \ll \Omega_1 $. It follows, that we can neglect
adiabatically the intrinsic dynamics of the oscillators $q_1$ and
$q_2$ and suppose it implicitly determined by the dynamics of the
squeezed level  polaron. (We confirmed numerically the instability
of the oscillations of $q_1$ and $q_2$). Hence, we can eliminate the
oscillators $ q_1 $ and $ q_2 $ from (\ref{6}) and  for the squeezed
level polaron $Q $, in the limit $j\rightarrow\infty$, i.e.
$\mu\rightarrow 0$, we obtain the dynamic nonlinear equation  for
the  classical "order parameter", 
\begin{equation}
\ddot{Q}=\frac{16\alpha^4}{\Omega_1^2}\left(1-\frac{\bar{\beta}^2}{\alpha^2}\right)Q+
\frac{3\cdot
2^4\alpha^4\bar{\beta}\sqrt{2(1-\bar{\beta^2}/\alpha^2)}}{\Omega_1^{5/2}}
Q^2 -\frac{2^6\alpha^6 (1-\bar{\beta}^2/\alpha^2)}{\Omega_1^3}Q^3\,,
\label{Q}
\end{equation}
or, using the transformation $Q=q+\displaystyle
{\frac{\bar{\beta}}{2\alpha}\frac{\Omega_1^{1/2}}{\sqrt{2(1-\bar{\beta}^2/\alpha^2)
}}}$ the quadratic term is eliminated and one obtains the normal
form
\begin{equation}
\ddot q=Aq-Bq^3+F,
 \label{q}
\end{equation}
where
\begin{equation}
A\equiv \frac{8\alpha^2(2\alpha^2+\bar{\beta}^2)}{\Omega_1^2}, \ \
B\equiv \frac{2^6\alpha^4(\alpha^2-\bar{\beta}^2)}{\Omega_1^3}>0, \
\ F\equiv \frac{4\alpha^2\bar{\beta}\sqrt{2
 (1-\frac{\bar{\beta}^2}{\alpha^2})}}{\Omega_1^{3/2}}+\frac{8\bar{\beta}^3}{\Omega_1^{3/2}\sqrt
{2(1-\frac{\bar{\beta}^2}{\alpha^2})}}. \label{par}
\end{equation}
 Contrary to the previous cases,  the transition described by equation (\ref{q}) between the normal
and radiation phase is of the first order and supports a coexistence
of both phases in the sector of the normal phase
$\alpha>\bar{\beta}$, or $\omega_{1}>\omega_{2}$. We come to the
conclusion that in contrast with the Dicke model the presence of the
additional oscillator opens a sector of a mixed phase with partial
occupation of the excitation space of all three coupled oscillators.
A transition between two non-equivalent minima of the potential
$V={\displaystyle \frac{B}{4}}q^4-{\displaystyle \frac{A}{2}}q^2-Fq$
arises and can be considered as the oscillation-assisted tunneling
(hopping). The formation of the radiation phase is driven by the
force $F$ which is compensated by the nonlinearity so that the
"nuclei" (bubbles) of the radiation phase stabilize for the set of
parameters (\ref{par}). The exact solution to (\ref{q}) is given by
\begin{equation}
q= a\frac{n_1+\cos (w \tau)}{n_2+\cos (w\tau)}, \label{w}
\end{equation} where
\begin{equation}
n_1=(2-n_2^2)/n_2, \ \ a^2= \frac{A}{B}\frac{n_2^2}{2+n_2^2}, \ \
w^2=2A\frac{n_2^2-1}{n_2^2+2}, \ \  a=-\frac{F}{2A}(2+n_2^2).
\label{fluc}
\end{equation}
The solution (\ref{w}) represents nonlinear non-sinusoidal periodic
oscillations. From equation (\ref{Q}) it is evident, that the
nucleation process  vanishes identically when approaching the
transition point $\omega_{1}=\omega_{2}$, where the effective
polaron localization energies $\omega_{1}$ and $\omega_{2}$ in both
sectors are equal. Let us remind that the scenario described above
can be perturbed by the fluctuations neglected by the adiabatic
elimination of other two oscillators ($q_1$ and $q_2$) and by
excitations due to the finite-size effects.

Let us note that, in spite of independence of  Eq. (\ref{Q}) on $j$
in the thermodynamic limit $j\rightarrow\infty$,  the dynamic
variable $Q$ is $j-$dependent so that
the resulting mixed phase exists only for finite $j$  as expected.

Numerical evaluation of the order parameter $J_z $ from the exact
Hamiltonian (\ref{H:gener}) presented in Fig. \ref{boson-fig2} illustrates
the existence of the intermediate phase by non-zero occupation in
the neighborhood of the point $\alpha=\bar{\beta}$.

\section{Interplay of quantum fluctuations and symmetry breaking}

 A qualitatively new situation occurs in the nonlinear regime at finite $j$,
 when all three  oscillators couple via interplay of fluctuations due to
 finite j and of breaking the rotational symmetry  ($|\alpha-\bar\beta|>0$).

 As the next step,  we shall include in (\ref{H}) small terms  up to $O(1/j)$ to find the
effect of quantum  fluctuations. To elucidate the role of
fluctuations and the difference between one- and two-boson cases
 let us apply first
the method we will use in what follows to the Dicke model
(\ref{2}). We displace again the operators (see \cite{Emary:2003:b})
\begin{equation}
a^{\dag}= c^{\dag}+\sqrt{\alpha}, \ b^{\dag}= d^{\dag}-\sqrt{\delta}
\label{displ2}
\end{equation}
 to exclude linear terms from Hamiltonian (\ref{2}).
Here, $\alpha$ and $\delta$ are the  mean values of excitation
number  $\langle a^{\dag}a\rangle$ and  $\langle b^{\dag}b\rangle$
of related bosons $a$ and $b$, respectively. Putting the Ansatz
(\ref{displ2})  into (\ref{2}) and eliminating linear terms, one
obtains two known solutions \cite{Emary:2003:b}
\begin{align}
& 1.  \  \alpha=0,\
 \delta =0 , \ {\rm normal \ phase, \ "1"};  \label{Dsol1}\\
 & 2. \  \sqrt{\alpha}= \frac{\lambda\sqrt{2j}}{\Omega}\left (
1-\left(\frac{\omega_0\Omega}{4\lambda^2}\right)^2\right )^{1/2}, \ \
\sqrt {\delta}= \sqrt j \left (
1-\frac{\omega_0\Omega}{4\lambda^2}\right )^{1/2}, \ {\rm radiation
\ phase,\  "2"} \label{Dsol2}
\end{align}
which determine  displacements of the normal phase "1" with zero
number of macroscopically excited bosons, and  of the radiation
phase "2"  at $\lambda^2> \omega_0\Omega/4$ with both numbers $\sim
j$. Then, Hamiltonian (\ref{2}) for the normal phase is reproduced,
\begin{equation}
H_{1D}= \Omega(c^{\dag}c+1/2)+\omega_0 (d^{\dag
 }d-j)+\lambda(c^{\dag}+c)\left (d^{\dag}+d-
 \frac{d^{\dag 2}d+d^{\dag}d^2}{4j}\right ).
 \label{H1D}
\end{equation}
 Let us write Heisenberg equations for the phase 1,
\begin{align}
& i \dot c= \Omega c +\lambda \left(d^{\dag}+d-
 \frac{d^{\dag 2}d+d^{\dag}d^2}{4j}\right) \,,\label{HD1}\\
& i \dot d = \omega_0 d +\lambda (c^{\dag}+c)
-\frac{\lambda}{4j}(c^{\dag}+c)(2d^{\dag}d+d^2)\,. \label{HD2}
\end{align}
 Within the adiabatic approximation, we can  neglect the intrinsic dynamics
 of the level oscillator $d$, assuming plausibly
  $\Omega\ll \omega_0$. Thus,  we suppose  $d  \langle d^{\dag}-d\rangle /dt =0$ in
(\ref{HD2}) so that the oscillator can be eliminated from this
equation and put into (\ref{HD1}). Defining then operators $\hat
q\equiv {\displaystyle \frac{1}{\sqrt {2\Omega}}}(c^{\dag}+c)$ and $\hat\pi\equiv i
{\displaystyle \sqrt{\frac{\Omega}{2}}}(c^{\dag}-c)$, the dynamic equation for the
coordinate $q= \langle \hat q\rangle$  up to the order $1/j$ can be
written:
\begin{equation}
\ddot q = -\Omega^2\left (1-\frac{4\lambda^2}{\omega_0\Omega}\right
) q -\frac{8\lambda^4\Omega^2}{j\omega_0^3}q^3 +O(1/j^2)\,.
\label{dynD}
\end{equation}
If ${\displaystyle \frac{4\lambda^2}{\omega_0\Omega}}<1$, there
result oscillations with the squeezed frequency $\Omega_D=
\Omega\left
(1-{\displaystyle\frac{4\lambda^2}{\omega_0\Omega}}\right )^{1/2}$
about the displaced center.
 The quasiclassical potential corresponding to equation (\ref{dynD})
 yields
\begin{equation}
V= \frac{\Omega^2}{2} \left(1-\frac{4\lambda^2}{\omega_0\Omega
}\right ) q^2+  \frac{\lambda^4\Omega^2}{4j\omega_0^3}q^4+V_0.
\label{W}
\end{equation}

For each $j$, the potential (\ref{W}) implies a second order phase
transition at $\lambda^2_c =\omega_0\Omega/4$ which has been found
as the critical point of the phase transition between the normal and
super-radiant phase in the Dicke model \cite{Emary:2003:b}. For
$\lambda<\lambda_c$, the normal phase is recovered being stable
while at $\lambda>\lambda_c$ the radiation phase is stable.  Let us
note, that the system (\ref{W}) is globally stable for each finite
$j$ and the phase transition
 is supported by the quantum fluctuations $\sim 1/j$.
The method we present here is so verified as being able to reproduce
basic results known from the standard approach used before
\cite{Emary:2003:b} and applied for the present model in  previous
Section.

In what follows we shall perform analogous calculations for the
system with two boson modes (\ref{H1}). For this case,  Heisenberg
equations related to (\ref{H1}) for the phase 1 read as follows
\begin{align}
& i \dot{c_1}= \Omega_1 c_1+\frac{2\alpha}{\sqrt {2j}}d^{\dag}d,
 \qquad \qquad  i \dot{c_2}= \Omega_2 c_2+ \beta
(d^{\dag}+d)-\frac{\beta}{4j}(d^{\dag 2}d +d^{\dag}d^2 ),
\nonumber\\
& i \dot d=\frac{4\alpha^2}{\Omega_1}d
+\frac{2\alpha}{\sqrt{2j}}(c_1^{\dag}+ c_1)d+\beta
(c_2^{\dag}+c_2)-\frac{\beta}{4j}(c_2^{\dag}+c_2)
(2d^{\dag}d+d^2)\,. \label{DE1}
\end{align}

 For strong interaction, it is plausible to assume
$\omega_1=\dfrac{4\alpha^2}{\Omega_1}\gg\Omega_1,\  \Omega_2$ and,
consequently, to apply the adiabatic approximation neglecting
intrinsic dynamics of the level polaron $d$ of high frequency
$\omega_1$, $\dot d= 0$. Consequently, it can be eliminated but it
determines the dynamics of the slaved modes $q_1=
(2\Omega_1)^{-1/2}\langle c_1^{\dag}+ c_1\rangle,\ q_2=(2
\Omega_2)^{-1/2}\langle c_2^{\dag}+ c_2\rangle$ implicitely. Hence,
we use the stationary expression for $q=
(8\alpha^2/\Omega_1)^{-1/2}\langle d^{\dag}+ d\rangle$ from
(\ref{DE1}) and put it into remaining equations for $q_1$ and $q_2$.
Up to the lowest order term in $1/\sqrt j$ one obtains
\begin{align}
&\ddot q_1=  -\Omega_1^2 q_1 -\frac{\beta^2\Omega_1\sqrt
{\Omega_1}\Omega_2}{4\alpha^3 \sqrt j} q_2^2 \label{qq:1}\\
&\ddot{q_2}=  -\Omega_2^2\left
(1-\frac{\beta^2\,\Omega_1}{\alpha^2\,\Omega_2}\right) q_2-
\frac{\beta^2\Omega_1^2\sqrt{\Omega_1}}{\alpha^3 \sqrt{2 j}}q_1 q_2\,. \label{qq:2}
\end{align}
 We received the squeezed frequency of the oscillator $q_2$ which justifies
 again the use of the slaving principle for the oscillator $q_1$,
$\Omega_1^2\gg\Omega_1^2(1-\bar{\beta}^2/\alpha^2)$. Its dynamics is
then implicitly ordered by the dynamics of the oscillator $q_1$.
Then, similarly as in the previous case, we receive
\begin{equation}
\ddot{q_2}= -\frac{d V(q_2)}{d q_2}\,, \label{dV}
\end{equation}
where
\begin{equation}
V(q_2)=\frac{\Omega_2^2}{2}\left
(1-\frac{\beta^2\,\Omega_1}{\alpha^2\,\Omega_2} \right)q_2^2
+\frac{\beta^5\Omega_1^2\Omega_2}{16\alpha^6 j\sqrt{2}}q_2^4+V_0\,.
\label{V}
\end{equation}

Equation (\ref{V}) implies a continuous phase transition at
$\alpha=\bar{\beta}\equiv \beta\sqrt{\Omega_1/\Omega_2} $  to the
order $1/j$ for each $j$
 between  the normal  oscillators (\ref{qq:2}) with squeezed frequency
 $\bar{\omega_2}=\Omega _2\left(1-\bar{\beta}^2/\alpha^2\right)^{1/2}$, ($\alpha>\bar{\beta}$), and a
 nonlinear sector analogous to the super-radiant phase of the Dicke
 model ($\bar{\beta}> \alpha$). Note that the potential for the Dicke model (\ref{W}) and (\ref{V}) become identical
if we set $\omega_0\equiv 4\alpha^2/\Omega$ in equation (\ref{W}).

We conclude that there occurs a sequence of phase transitions for
each finite $j$ between the quasi-normal and radiation phase
($\bar\beta>\alpha$) because of softening (squeezing) the frequency
of the oscillator $\bar{\omega_2}=$ by the parameter
$\bar{\beta}/\alpha$.

According to Eq. (\ref{epscrit1}) and (\ref{epscrit2}) the critical
energies at both sides of the phase transitions are symmetric when
changing $\alpha\leftrightarrow \beta $ and $\Omega_1\leftrightarrow
\Omega_2$. In the radiation phase, for $j$ finite, the nonlinearity
bears one-instanton solution
\begin{equation}
\bar{q_{21}} (\tau-\tau_0)=
\pm\frac{2\alpha_1^3}{\bar{\beta}^2\Omega_1}\sqrt
{\frac{2j}{\Omega_1}\left(
1-\frac{\alpha^2}{\bar{\beta}^2}\right)}\tanh\left(\frac{\Omega_1}{\sqrt
2}\left
(1-\frac{\alpha^2}{\bar{\beta}^2}\right)^{1/2}(\tau-\tau_0)\right)\,.
\label{K1}
\end{equation}

Here, $\bar {q_2}= q_2\sqrt {\frac{\Omega_1}{\Omega_2}}, \ \bar
{\beta}= \beta\sqrt {\frac{\Omega_1}{\Omega_2}}$.  The one-instanton
solution (\ref{K1}) is associated with the tunneling between the
extrema of the potential inverted to (\ref{V}) if $\bar{\beta}>
\alpha$ and $\tau\rightarrow it$. Hence, at finite $j'$s  there
appears a new instanton phase as a precursor of the phase transition
at the maximum softening of the frequency at $\bar{\beta}
\rightarrow \alpha$ in the radiation phase. More generally, there
occurs a sequence of repeated tunnelings (oscillations between two
equivalent minima of a local potential) for each lattice site. The
same effect occurs in the Dicke model if
$4\lambda^2/(\omega_0\Omega) >1$ (\ref{W}). Moreover, in the present
model, there exists the coupling between the oscillators $1$ and $2$
(\ref{qq:1},\ref{qq:2}) which was neglected in the adiabatic
approximation (\ref{dV}). In fact, within more subtle calculations
there would occur tunnelings mediated by two coupled oscillators
(one of them being a polaron) for each $j$ instead of one of the
adiabatic treatment.

 The time of the tunneling $T=\omega_T^{-1}$ is
determined by the squeezing of the frequency $\omega_T=
\dfrac{\Omega_1}{\sqrt
2}\left(1-\dfrac{\alpha^2\Omega_2}{{\beta\Omega_1}^2}\right
)^{1/2}$, i.e. the tunneling ceases at the transition point. The
frequency $\omega_T$ determines the curvature close to the minima of
the bistable potential (\ref{V}). This initial frequency  enters the
probability of the tunneling from the ground state of  the
oscillator close to the minimum during the time $\tau $ when
starting at $\tau_0$ (e.g., \cite{Dekker}). According to Dekker
\cite{Dekker} the trajectory of the tunneling particle in time
exhibits chaotic features if the initial frequency is squeezed. This
implies that the quantum tunneling should be associated with the
chaotic features inherent to finite lattice or excited spectra of
the standard Dicke model (\ref{W}) or our model in the present
approximation. However, there is self-evident a question about the
possible mechanism of chaos production due to the squeezing in this
models. Going from the present considerations, instead of the strict
tunneling (\ref{K1}), we propose a more realistic mechanism of
hopping, i.e., the above mentioned tunneling between the squeezed
minima assisted by the neglected (when going from (\ref{DE1}) to
(\ref{dV})) nonadiabatic fluctuations of the oscillators $c_1$ and
$d$ coupled by the nonadiabatic nonlinear terms $\sim 1/\sqrt j$ and
similarly of the oscillator $d$ (\ref{HD2}) for the Dicke model.
 We remark in passing that the mechanism of hopping is known, e.g., from the transport theory of disordered
 systems at finite temperatures \cite{Overhof}.

\section{Statistical characteristics  of excited levels and wave functions} \label{stat}

The distribution of the nearest neighbor level spacings (NNS) of
excited quantum levels is the standard testing point for
investigation of the issues in quantum chaology \cite{Dyson}.
Recently we have reported results on NNS distribution of the
 molecular JT model (\ref{HJT}) \cite{Majernikova:2006:b}
which is the case $j=1/2$, $\Omega_1=\Omega_2$, $\alpha\neq\beta$ in
the notation of the present article.
 It was shown that these distributions essentially deviate from the Wigner
distribution typical for well developed quantum chaos. Our results
show NNS distributions nonuniversal but far from the cases with
higher symmetry (rotational $\alpha=\beta$ and linear $\alpha
\gg\beta$, $\alpha\ll\beta$) they tend to a form close to the
semi-Poisson distribution $P(S)= 4S \exp (-2S)$ already encountered
in some problems, in particular in the Anderson model close to the
metal-insulator transition \cite{Shklovski}. The peculiarity of the
Jahn-Teller system with equal frequencies was that the Wigner-Dyson
level spacing distribution has never been reached, and the said
semi-Poisson form of the statistics appeared to be the most
``chaotic'' one.

In this section we present statistical characteristics (NNS
distributions, entropy of level occupation, and spectral density of
states) of the spectra for $\Omega_1\neq\Omega_2$, $j\geq 1/2$. We
solved numerically the eigenvalue problem for the quantum
Hamiltonian ({\ref{H:gener}). The Hamiltonian was diagonalized with
the basis of the  boson Fock states for bosons $1$ and $2$. When
taken $N_1$ and $N_2$ Fock states for each of $2j+1$ electron
levels, there is produced an inevitable truncation error, so that
the results were checked against the convergence (with changing the
numbers $N_1, N_2$). Only about $\sim 1100$ lower states were used
for calculating the statistics out of typically $\sim (8\div 10)
\times 10^3$. The obtained raw energy spectrum had to be treated by
an unfolding procedure in order to ensure the homogeneity of the
spectrum (constant local density of levels). Thereafter the
statistical data were gathered in a standard fashion from small
intervals in the space of parameters $(\alpha, \beta)$. The
calculations were performed essentially along the same lines as our
previous calculations for the Jahn-Teller problem with $j=1/2$
\cite{Majernikova:2008,Majernikova:2006:b} where additional details
were given as to the convergence check, unfolding procedure and
gathering statistics.

 The results for the level spacing distributions for different phonon
frequencies $\Omega_1\neq \Omega_2$ and $j>1/2$ show rather vast
variety ranging from Wigner to Poisson distributions as limiting
cases but recovering also the semi-Poisson distribution in a rather
wide range of model parameters. With increasing $j$, general
tendency consists in increasing the diapason of parameters where the
NNS shows deviations from the Poisson distribution towards
chaoticity up to the maximum degree of chaoticity given by the
Wigner distribution. So, for the case $j=1/2$ and equal frequencies
 the maximal degree of chaoticity being expressed by the semi-Poisson
distribution in accord with our previous results. For moderate $j$
($j=7/2$ and $\Omega_1\neq \Omega_2$ exemplified in \ref{boson-fig3}
there appears a well-marked domain of maximal chaoticity (Wigner
distribution). For example, for $\Omega_2/\Omega_1=2$ the domain of
Wigner distribution stretches for the values of parameters
$1\lesssim \alpha,\beta\lesssim 3$.

We can conclude that the presented results essentially deviate from
the Wigner statistics of levels imposed by the one-boson Dicke model
\cite{Emary:2003:b}. For comparison, we performed the same
calculations of NNS distributions for the standard one-boson Dicke
model  (Fig. \ref{boson-fig4}). The distributions follow closely the
Wigner form of the $P(S)$ curve, but a good agreement is achieved
only for high values of $j$.  For the resonance case
$\Omega_1=\Omega_2$  the quantum chaotic statistics (Wigner surmise
\cite{Dyson}) changes towards a non-universal with respect to
$\alpha$, $\beta$ but tending to the semi-Poisson intermediate
statistics achieved asymptotically far from the cases of the special
symmetry (rotation, linear).
 In contrast to one-boson models, this universality is observed
 even for small number of electron levels and seems not to change with raising
$j$. Therefore the impact of the second boson mode is that it
essentially changes the quantum statistics. For the non-resonance
case $\Omega_1\neq\Omega_2$ we observe a considerable suppression of
chaos manifesting in the reduction of the ``pure'' domain of Wigner
chaos in the space of the parameters $\alpha,\beta$. 

 To visualize
the wave functions of the excited levels of the system we
 considered the spreading of the wave functions over the electron
levels and integrated out two boson degrees of freedom. Thus we
found numerically the probability $P_i^{(n)}$ of the wave
function $\phi_i^{(n)}$ to occupy a given electronic level $i$ of
the system. The full wavefunction depending on both vibron
and electron level variables is calculated by the same scheme of the
diagonalization of the quantum Hamiltonian in the representation of
Fock states as described above. If $\beta=0$, Hamiltonian equation
is split onto $N(=2j+1)$ independent equations labeled by
$i=1,\dots,(2j+1)$, so that the wavefunction for each given state
$n$ is localized on some electron level $i$. If $\beta\neq 0$,
Hamiltonian matrix is no more diagonal (since $J_x$ is nondiagonal), so
that its components get interacting. The resulting eigenfunction is
in general $(2j+1)$-fold function occupying each electronic level
(see Chapter 1 as for the discussion about normal and super-radiant
phase in the thermodynamic limit). Let $\chi_i^{(n)}(Q_1,Q_2)$ be
the i-th component of the $(2j+1)$-dimensional vector of the
eigensolution of the Hamiltonian matrix equation for n-th energy
level in the "coordinate" representation ($\hat{Q}_l\propto
b_l^{\dag} +b_l$, l=1,2). Then the
  occupation probabilities  of the i-th electronic level are $P_i^{(n)}=\iint
|\chi_i^{(n)}|^2 \mathrm{d} Q_1 \mathrm{d} Q_2$. In this
representation  we exemplify the wave functions in Fig. \ref{boson-fig5}
for the levels $n=1,4$ and the parameters around the point of QPT
$\alpha=2$,  $\beta=1.95, 2.05$ and $\Omega_1=\Omega_2$. The abrupt
change in the shape of wavefunctions when going from normal to
super-radiant phase is easily perceivable not only for the ground
state, but for lowest excited states as well (in this example for
n=4). However, for higher excited states the wavefunctions generally
spread over the whole available electron space, irrespective to the
relation between $\alpha$ and $\beta$. The last relation  is also
reflected in the statistical properties of levels: they are
invariant with respect to the  exchange $\alpha \leftrightarrow
\beta$, $\Omega_1\leftrightarrow \Omega_2$, likewise it was already
shown for the generalized JT model \cite{Majernikova:2006:b}. In
order to characterize quantitatively the behavior of the
wavefunctions of excited states we introduce the entropy of level
occupation
\begin{equation}
S_n= - \sum \limits_{i=1}^N P_i^{(n)} \log P_i^{(n)} \,,
 \label{Sn}
\end{equation}
where $\sum\limits_{i=1}^N P_i^{(n)} =1$. In Fig. \ref{boson-fig6}
for $\Omega_1=\Omega_2 $ we plot the said entropies for first 1000
quantum levels of a system with four electron levels ($j=3/2$). One
can see from these figures that in the most excited states  the
electronic levels are equally populated (with probabilities
$1/(2j+1)$, e.g., last Figure in Fig. (\ref{boson-fig5})),  thus the
said entropies tend to their limiting value $\log (2j+1)$. For four
levels the entropy yields the value $S_4 = 2\log 2= 1.38$ which is
confirmed by Fig. \ref{boson-fig6} for $\alpha\neq \beta$. However,
among these extended states there emerge ``localized'' levels of
lower values of the entropy which are characterized by a relative
localization of wavefunction on smaller number of electron levels.
Such levels form marked branches seen in Fig. \ref{boson-fig6},
which remind of the branches of ``exotic'' states in the generalized
Jahn-Teller model \cite{Majernikova:2006:a}. Most of the states show
an intermediate degree of localization between both of the limits.
These intermediate electronic states correspond to the mixed
intermediate domain of the normal and radiation phase in the
respective boson space.

Another way to  visualize the complex structure of the excited
states and corresponding wave functions is the representation of the
spectral density given by the imaginary part of the projected
resolvent \cite{Ziegler:2005}.  It shows the characteristic
frequencies of the final state of the evolution of a system starting
from the projected Hilbert space and returning to it. The spectral
density of states $F(E)$ can be defined with respect to some initial
state $|\Psi_0\rangle$ (the ground state, see below) and is
connected with the return probability to this state in the course of
the system evolution through the exact states $\Phi_n$,
eigenfunctions of the energies $E_n$ of the system at small
$\varepsilon \sim 0$:

\begin{equation}
F(E)\equiv \mathrm{Im}
 \langle \Psi_0 |(E-\hat{H}-i\varepsilon)^{-1}
|\Psi_0\rangle = \varepsilon\sum_n \frac{|\langle
\Phi_n|\Psi_0\rangle |^2}{(E-E_n)^2+\varepsilon^2}, \label{spectfun}
\end{equation}
where $|\Phi_n\rangle$ and $E_n$ are the eigenfunctions and
corresponding eigenvalues of the Hamiltonian $\hat{H}$. The small
parameter $\varepsilon$ fixes the rules of handling the poles of the
Green function in the complex space. In Fig. (\ref{boson-fig7}) we
show the examples of the spectral density for the values of
parameters $\alpha$, $\beta$ below and above the critical line. As
the initial state $|\Psi_0\rangle$ we took the ground state with
$\alpha=\beta=0$, that is the product of the electronic Dicke state
$|j,-j\rangle_{el}$ and the phonon state $|0,0\rangle_{ph}$ with
zero number of both bosons 1 and 2. Thus the spectral density in
Fig. (\ref{boson-fig7}) characterizes the evolution of the system
with the interactions $\alpha$, $\beta$ switched on in the initial
time. The peaks at each $E_n$ measure the overlap between the
initial state $\Psi_0$ and the eigenstate $\Phi_n$ for
$\varepsilon\sim  0$.

\section{Conclusion}

The numerical analysis of the generalized (two boson modes with
different frequencies) Dicke model or the equivalent generalized
lattice JT model with fully broken local rotation symmetry
(different coupling constants $\alpha\neq\beta$ and different
frequencies $\Omega_1\neq\Omega_2$) brings following new results:

 (i) The critical point of the second order phase transition , (\ref{boson-fig1}), is
modified to $\alpha=\beta \sqrt{\frac{\Omega_1}{\Omega_2}}$ when
compared with the critical point of  the one-boson Dicke model
$\alpha=\beta$. The renormalization of the couplings by the
frequencies is due a reformulation of the critical point in terms of
 frequencies of the dressed oscillators as $\omega_1=\omega_2$, $\omega_1=\frac{4\alpha^2}{\Omega_1}$
, $\omega_2=\frac{4\beta^2}{\Omega_2}$.

(ii) The electron level order parameter $\langle J_z\rangle/j$
(\ref{boson-fig2}) shows  the finite size effect for finite $j$
close to the phase transition point which smoothes the phase
transition to a rather broad at small $j$ "mixed" region. The effect
is slightly nonsymmetric around the critical point supporting the
order parameter in the normal phase.

 (iii) The level spacing probability distribution (\ref{boson-fig3})of the model
under consideration bears a drastic change from the Wigner-Dyson
$P(S)\sim S\exp(-S^2)$ distribution of the one-boson Dicke model
(\ref{boson-fig4}) by a non-universal transition up to close to the
Poisson distribution $P(S) \sim \exp (-S)$. For equal frequencies,
$\Omega_1=\Omega_2$ and different coupling constants, however,  the
distribution ranges non-universally from a close to the semi-Poisson
$P(S) = 4S\exp (-2S)$ to  the Poisson distribution. Evidently, the
change to the range between the Wigner-Dyson and the semi-Poisson
distribution is due to the difference in the frequencies.

(iii) In the electron space, the phase transition reveals in a
drastic change of the wave functions from those spread over small
number of levels (sites) in the "normal"  phase to the macroscopic
distribution of the excitation over all the levels (lattice) in the
radiation phase, (\ref{boson-fig5}). This behavior shows up a
non-universality with regard to the interaction parameters and
inhomogeneity throughout the spectral levels.

(iv) Entropies of occupation of the electronic levels
(\ref{boson-fig5}) illustrate the measure of localization of the
states throughout the spectra (or the lattice). There is evident the
asymptotic line of the delocalized states at $S=1.38$ and the
localized states with various extent of localization. Close to the
line from below there is evident a relatively high density of weakly
correlated states. The lowest levels show up branches of the
localized states. Let us remind that the exceptional case of
$\alpha=\beta$ belongs to a higher class of symmetry out of the
concept of the present paper.

(v) A measure of the localization of the spectral states as a
function of energy is the spectral density of states
(\ref{boson-fig7}). It measures the evolution of the states from the
pure Dicke states ($\alpha=\beta$) in $t=0$ to the exact numerical
states of the system at interactions switched on at $t>0$. During
the evolution of the system to the final state the system overlaps
with the initial state at the poles of respective propagator $F(E) $
(\ref{spectfun}). The peaks of the function $F(E) $ measure the
overlap between the initial Dicke states $\Psi_0$ and the exact
eigenstates $\Phi_n (E)$ for $\varepsilon\sim 0$.

Beside the exact numerical analysis, use of the Holstein-Primakoff
Ansatz made us possible to treat analytically, though approximately,
the present model. The analytical treatment brings a useful insight
into the mechanisms behind of the properties illustrated by the
numerical results. We have shown that an additional boson mode
coupled with the electron level spacing dresses the normal-phase
oscillator to a self-trapped one similarly as it was dressed the
oscillator in the radiation phase. This mode-dressing modified by
squeezing cooperates with nonlinear terms at finite $j$:  It
modifies both the normal and the radiation phase close to the
transition point $\alpha=\beta \sqrt{\dfrac{\Omega_1}{\Omega_2}}$.
The intermediate region of the "nucleation"  occurs as a consequence
of the competition of the self-localization due to the polaron
dressing modified by squeezing of the level boson mode and of the
boson-assisted tunneling. Both quasi-normal and radiation phases
coexist within the sector of the quasi-normal phase
$\alpha>\bar{\beta}$. We have shown that the formation of the
radiation phase is driven by the boson-assisted tunneling (hopping)
and is compensated by the nonlinearity due to the selftrapping so
that the "nuclei" of the radiation phase get stabilized.  So, in
contrast with the one-boson Dicke model, for finite $j$ the
additional oscillator opens a sector of a "quantum mixed phase" with
partial occupation of the excitation space of all three coupled
oscillators.

  We have shown that in
the radiation phase there occurs a sequence of repeated tunnelings
(oscillations between two equivalent minima of a local potential)
 for each lattice site. The radiation phase can be then
 represented as an almost ideal instanton--anti-instanton gas phase (instanton lattice).
However, there exists a weak coupling with two oscillations
assisting the tunneling because of non-adiabatic fluctuations due to
the finiteness of the lattice. They are necessarily involved by the
model and moderate the polaron dressing of the level mode
(\ref{H3}). As a result, they perturb the ideal
periodicity of the instanton lattice. In a variable extent the
quantum fluctuations due to the level correlations persist within
the whole radiation phase and represent the dephasing mechanism of
thorough coherence of the phase.
 We have shown that the sequence of local "second-order phase transitions"
 and so the instanton--anti-instanton gas phase exists in the standard one-boson
Dicke model as well, in contrast to the mixed phase which does not
exist in this model.

Numerical calculations show reflection of the described three phases
in the statistical properties. We have demonstrated a drastic
qualitative change of the level spacing probability distributions
from the universal Wigner of the one-boson Dicke model to the
non-universal transition from the Wigner distribution of the
localized states of the quasi-normal phase through the critical
semi-Poisson distribution of the mixed domain to a close-to-Poisson
distribution for the almost ordered radiation phase. This change is
attributed to the weakening of the Wigner chaos caused by the
tunneling mechanism mediated by the additional mode.   The
fluctuation effects in the "radiation" phase are respectively weakly
$j$-dependent for small $j$.

For finite lattice systems, where the localization-delocalization
phase transitions occur, we suggest possibility to apply the present
results to interpret finite-size fluctuation effects. We propose the
applications of the presented results for finite systems where
fluctuations destroy quantum coherence and a second order phase
transition is modified by the tunneling mediated by the
fluctuations. The destroying the coherence of the radiation phase by
 the boson selftrapping is a source of the dephasing
mechanism in the radiation phase. The spectral phenomena as is the
inhomogeneous broadening of the zero-phonon spectral lines and the
broadening of spectra are due to the selftrapping in the
quasi-normal phase. This interpretation can be applied, e.g., as a
mechanism responsible for the inhomogeneous broadening of the
zero-phonon lines of the Dicke superfluorescence in KCl
\cite{Malcuit:1987}.

 The finite-size fluctuations enable the formation of a critical phase of
mixed localized and delocalized phases. An example of such a system
 with analogous picture of quantum phase transition provides the ultracold Bose gas in an optical
 lattice where the  phase coherence of Bose atoms is destroyed by the suppression of the tunneling
 due to the localization by the strong lattice potential \cite{Ziegler:2005}, \cite{Roth:2004}.
 From the competition of the tunneling and the localization due to the lattice
 there results the phase transition between the highly coherent superfluid phase of Bose atoms and
  the localized Mott insulating phase of the atoms on the lattice
  \cite{Fisher:1989}. If some fluctuations, e.g., due to lattice vibrations, mediated the tunneling
  within the insulating phase, as well as perturbed the coherence of
  the coherent superfluid phase, the system of Bose atoms in the optical lattice would exhibit analogous properties
  as those of the two-boson Dicke model.

 We propose also the additional mode as a mean to reduce chaos in related spectral
properties of optical devices based on the Dicke model.

 {\bf Acknowledgement} The
financial support by the project No. 2/0095/09 of the Grant Agency
VEGA of the Slovak Academy of Sciences
 is highly acknowledged.

\begin{figure}[htb]
\includegraphics[width=0.6\hsize]{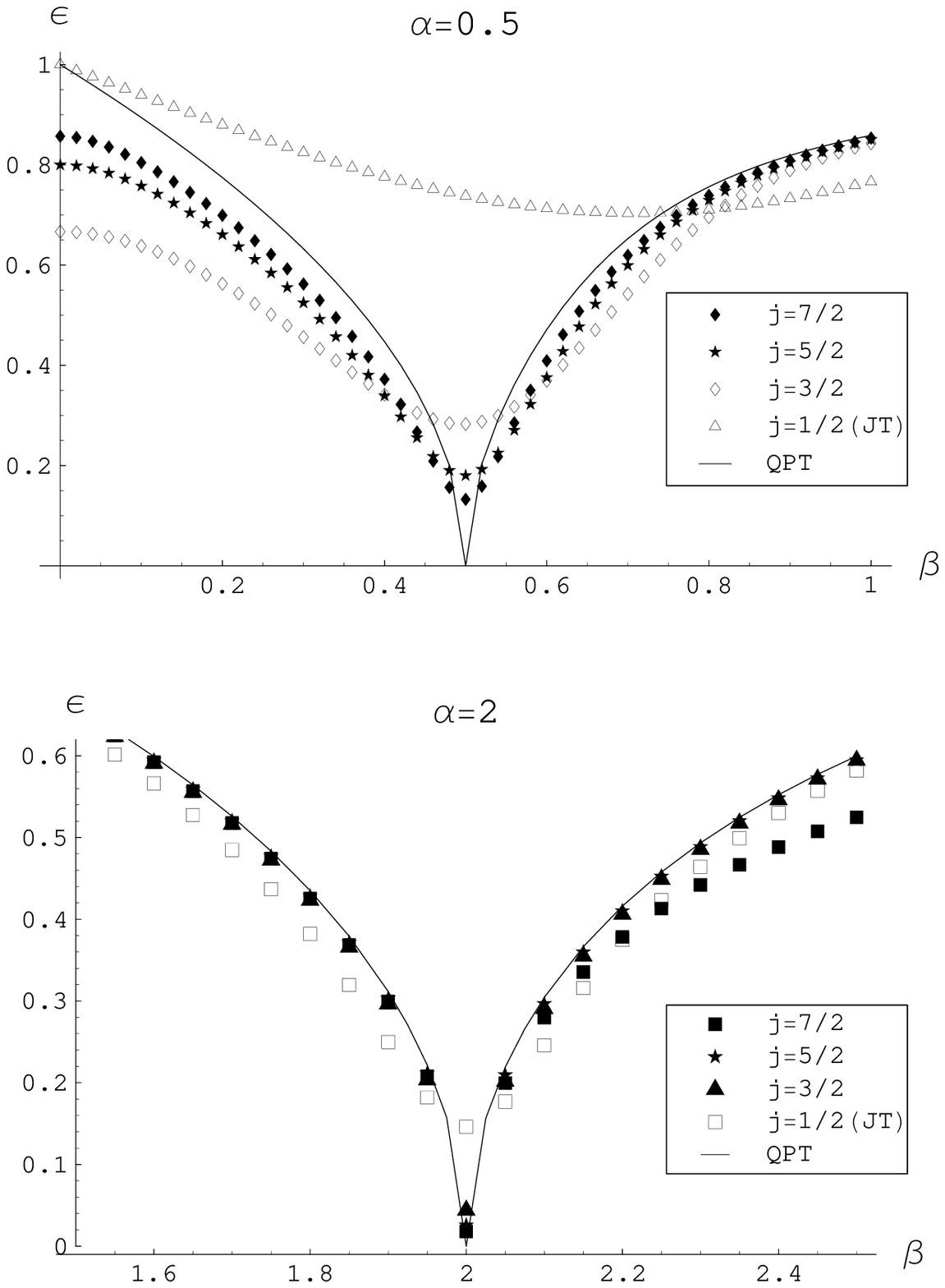}
\caption{ Phase transition between the normal and radiation phase in
the generalized Dicke and J-T lattice model for $\alpha=0.5$, $2$
and resonance case $\Omega_1=\Omega_2=1$. The symbols show the
numerical results for the excitation energy of the first excited
state for different $j$, the solid line in each figure is the
analytical result for QPT (\ref{eps1}), (\ref{eps2}) valid for
$j\to\infty$ . The cusp-like behavior at the critical point
$\omega_1\equiv4\alpha^2/\Omega_1=4\beta^2/\Omega_2^2\equiv\omega_2$
appears already for relatively small number of sites, e.g.,  for
$j=7/2$. For small $j$, the fluctuations smooth the cusp especially
at weak couplings. The non-symmetry about the critical point by the
reduction of the energy in the radiation phase for $\alpha=0.5$ is
due to the effect of quantum fluctuations (squeezing) and vanishes
with increasing $j$ and $\alpha$. For the case of different
$\Omega_1\neq\Omega_2$ the picture is qualitatively the same, there
occurs a shift of the transition point inwards the phase with higher
$\Omega_i$.} \label{boson-fig1}
\end{figure}

\begin{figure}[htb]
\includegraphics[width=0.7\hsize]{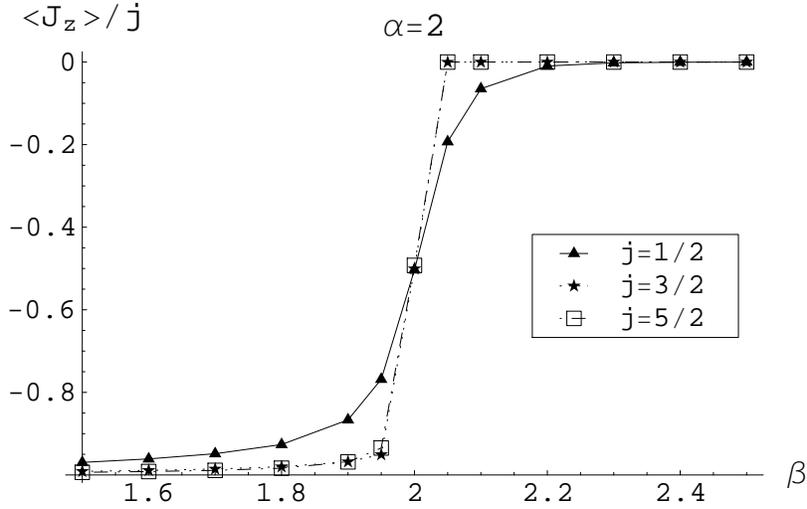}
\caption{  Order parameter $\langle J_z\rangle/j$ for $j=1/2, 3/2,
5/2$ and $\Omega_1=\Omega_2$. The finite-size effect ($j$ finite) of
the mixed phase about the transition vanishes for $j\rightarrow
\infty$.}\label{boson-fig2}
 \end{figure}

\begin{figure}[htb]
\includegraphics[width=0.8\hsize]{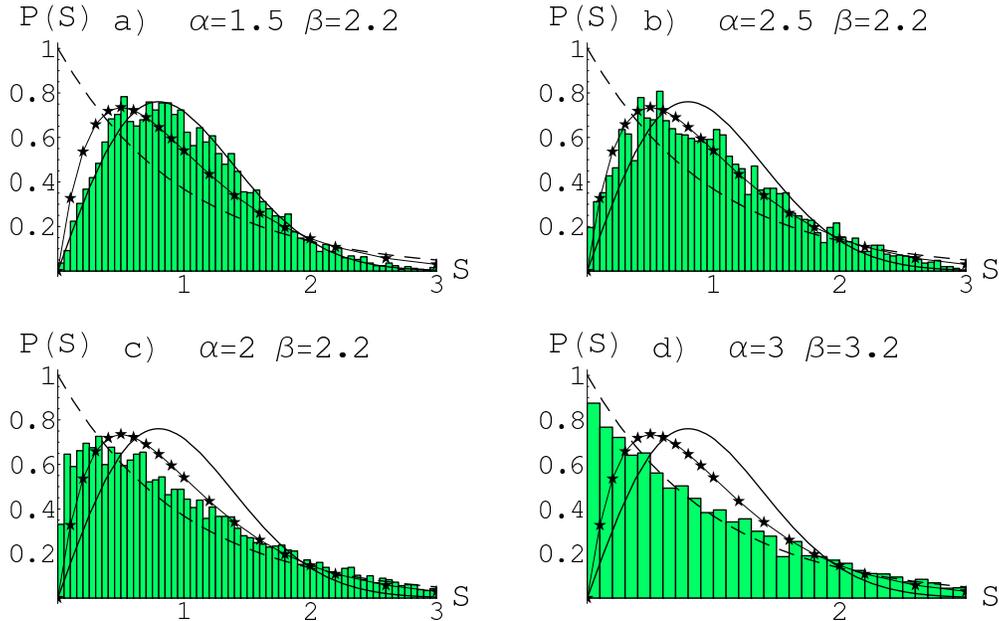}
\caption{ Level spacing distributions for $j=7/2$ and different
($\alpha$, $\beta$) (parameters are scaled to $\Omega_1=1$). Upper
row ((a,b)): $\Omega_2= 2\,\Omega_1$; bottom row ((c,d)): $\Omega_2=
0.5\,\Omega_1$. The curves show Wigner (solid), Poisson (dashed) and
semiPoisson (stars) distributions.
 NNS distributions in b), c) are close to the semi-Poisson distribution $P(S)=4 S \exp (-2S)$
\cite{Majernikova:2006:b}; histograms in a) and d) are almost
perfect Wigner ($P(S)\sim S\exp(-S^2)$ and Poisson $P(S) = \exp
(-S)$ distributions.} \label{boson-fig3}
\end{figure}

\begin{figure}[htb]
\includegraphics[width=0.7\hsize] {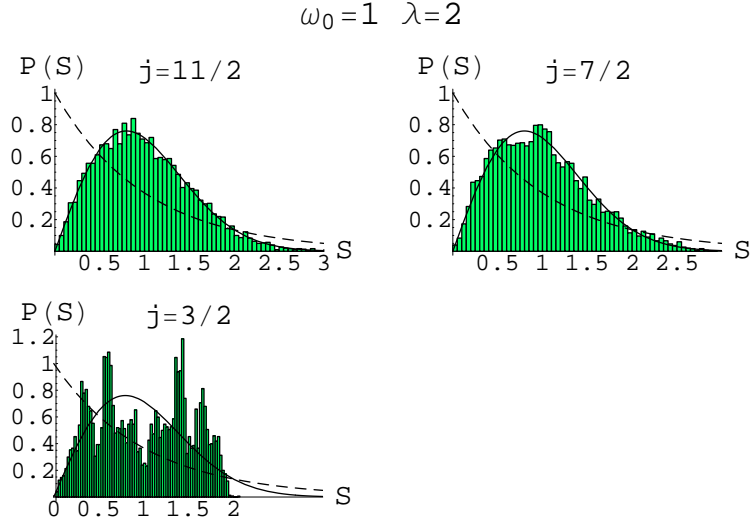}
\caption{ Level spacing distributions for one-boson Dicke model (\ref{2})
(boson frequency $\Omega$ scaled to $1$). At
large $j=7/2, 11/2$ almost pure Wigner chaos $P(S)\sim S\exp (-S^2)$
is revealed. The smaller $j$, the stronger are level fluctuations
growing with decreasing $\lambda$. } \label{boson-fig4}
\end{figure}

\begin{figure}[htb]
\includegraphics[width=0.8\hsize]{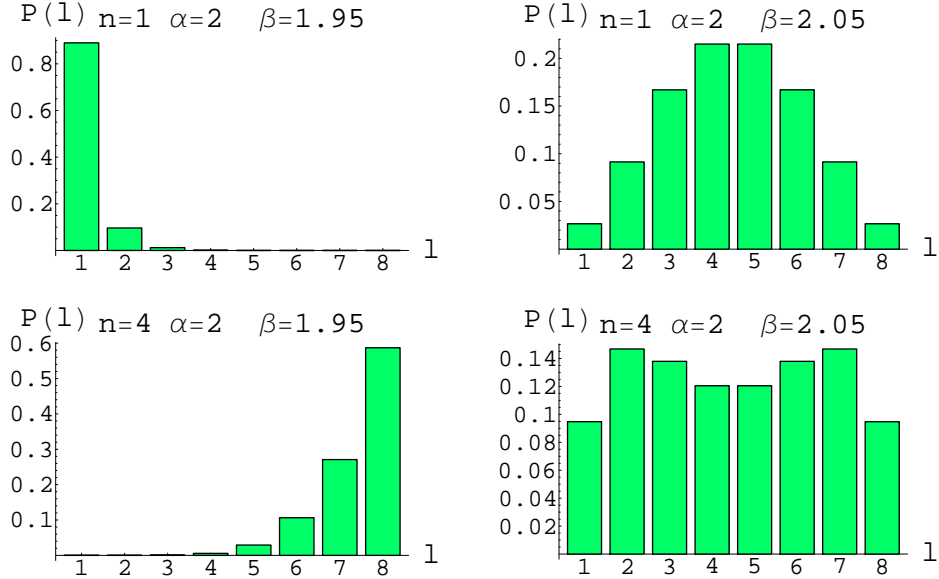}
\caption{The occupations $P(l)$ of the electronic levels $l= 1,\dots
2j+1$ (see Sect. IV.) close the the critical point $\alpha=\beta$
($\Omega_1=\Omega_2=1$; $j=7/2$) of the phase transition between the
quasi-normal and radiation phase for the ground $n=1$ and fourth
$n=4$ excited state. For $\Omega_1\neq\Omega_2$ the occupations
remain qualitatively the same. } \label{boson-fig5}
\end{figure}

\begin{figure}[htb]
\includegraphics[width=0.8\hsize]{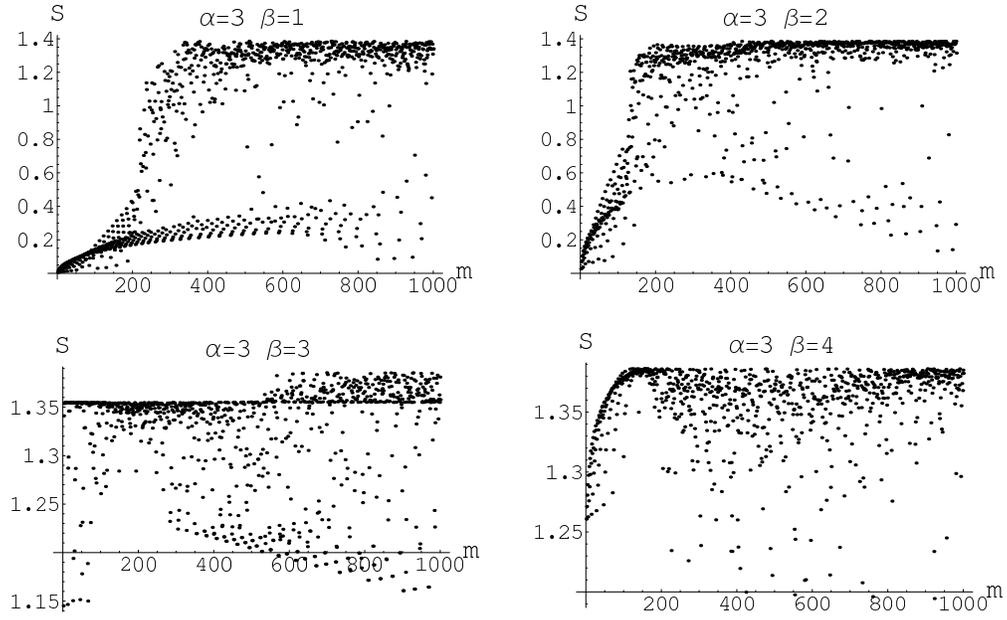}
\caption{Entropies of occupation of electronic levels (\ref{Sn}) as
function of the number of excited state $m$ for $j=3/2$ (4 levels).
The states with entropy lower than the limiting value $\log (2j+1)$
have larger measure of localization.   } \label{boson-fig6}
\end{figure}

\begin{figure}[htb]
\includegraphics[width=0.8\hsize]{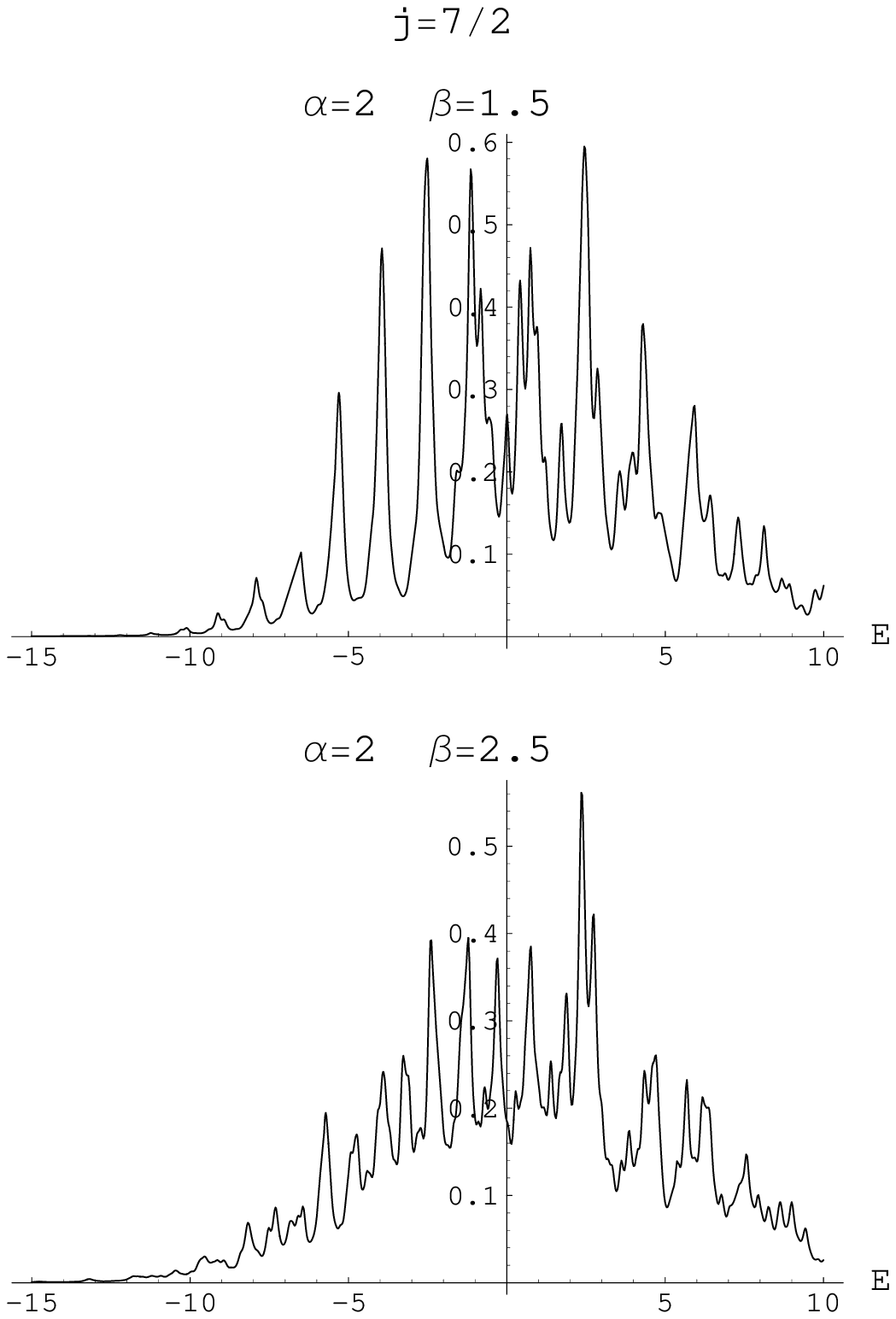}
\caption{Examples of the spectral density of states $F(E)$
(\ref{spectfun}) for $\alpha=2$ and $\beta=1.5, 2.5$ (We plot the
resonant case $\Omega_1=\Omega_2=1$; $j=7/2$; the non-resonant cases
do not change the picture qualitatively. The same applies for
previous two Figs.) } \label{boson-fig7}
\end{figure}


\begin{thebibliography}{99}
\bibitem{Graham:1984:a} P. Graham, and M. H\"{o}hnerbach, Z. Phys. B {\bf 57}, 233
(1984).
\bibitem{Graham:1984:b} P. Graham, and  M. H\"{o}hnerbach, Phys. Lett. A {\bf 101A}, 61
(1984).
\bibitem{Lew:1991} C.H. Lewenkopf,  M.C. Nemes, V. Marvulle, M.P.
Pato, and W.F. Wreszinski, Phys. Lett. A {\bf 155} 113 (1991).
\bibitem{Cibils:1995} M. Cibils, Y. Cuche, and G. M\"uller, Z. Phys. B {\bf 97},
565 (1995).
\bibitem{Dicke} R. H. Dicke, Phys. Rev. {\bf 93}, 99 (1954).
\bibitem{Emary:2003:a} C. Emary, T. Brandes, Phys. Rev. Lett. {\bf 90}, 044101
(2003).
\bibitem{Emary:2003:b} C. Emary, T. Brandes, Phys. Rev. E{\bf 67}, 066203
(2003).
\bibitem{Haake} K.-D. Harms, and F. Haake, Z. Phys. B {\bf 79}, 159 (1990);
S. Gnutzmann, F. Haake, and M. Kus, J. Phys. A {\bf 33}, 143 (2000).
\bibitem{Tolkunov:2007} D. Tolkunov, and D. Solenov, Phys. Rev. B {\bf 75}, 024402
(2007).
\bibitem{Pfeifer} P. Pfeifer, Phys. Rev. A {\bf 26}, 701 (1982).
\bibitem{Larson:2008} J. Larson, Phys. Rev. A {\bf 78}, 033833 (2008).
\bibitem{Skribanowitz:1973} N. Skribanowitz, I. P. Herman, J.C.
MacGilliwray, and M. S. Feld, Phys. Rev. Lett., {\bf 30}, 309
(1973).
\bibitem{Gross:1982} M. Gross, and S. Haroche,  Physics Reports {\bf 93}, 301
(1982).
\bibitem{Andreev:1980} A. V. Andreev, V. I. Emel'yanov, and Yu. A.
Il'inskii', Sov. Phys. Usp. 23, 493 (1980).
\bibitem{Florian:1984} R. Florian, L.O. Schwan, and D. Schmid, Phys.
Rev. A {\bf 29}, 2709 (1984).
\bibitem{deVoe:1996} R. G. DeVoe,  and R. G. Brewer, Phys.Rev.Lett {\bf
76} 2049 (1996).
\bibitem{Brandes:2000} T. Brandes, and B. Kramer,  Physica B {\bf
284} 1774 (2000).
\bibitem{Brandes:2005} T. Brandes, Phys. Repts, 408, 315-474 (2005).
\bibitem{Jarett:2007} T.C. Jarett, A. Olaya-Castro, and N. F. Johnson,
J. Phys. Conference Series, {\bf 84}, 012009 (2007).
\bibitem{Scully:2008} M. O. Scully, Y. Rostovtsev, A. Svidzinsky, and J-T.
Chang, J. Mod.Optics {\bf 55}, 3219 (2008).
\bibitem{Kuusmann:1975} I.L. Kuusmann, P.K. Liblik, and Ch. B.
Lushchik, JETP Lett., 21, 72 (1975).
\bibitem{Kmiecik:1987} H. J. Kmiecik, M. Schreiber, T. Kloiber, M.
Kruse, and G. Zimmerer, J. Lumin. {\bf 38}, 93 (1987).
\bibitem{Kishigami:1992} T. Kishigami-Tsujibayashi, and K. Toyoda,
T. Hayashi,  Phys. Rev. B {\bf 45}, 13 737 (1992).
\bibitem{Ding:1997} Xiaoya Ding, and John C. Wright, Chem. Phys. Lett.
269, 341 (1997).
\bibitem{Malcuit:1987} M. S. Malcuit, J. J. Maki, D. J. Simkin, and
R. W. Boyd, Phys. Rev. Lett. {\bf 59}, 1189 (1987).
\bibitem{Kongeter:1990} A. K\"ongeter and M. Wagner, J. Chem. Phys.
{\bf 92}, 4003 (1990); A. K\"ongeter and M. Wagner, J. Lumin. {\bf
45}, 235 (1990).
\bibitem{Eiermann:1992} H. Eiermann and M. Wagner, J. Chem. Phys. {\bf 96}, 4509
{1992}.
\bibitem{Slonczewski} J. C. Slonczewski, Phys. Rev. {\bf 131}, 1596
(1963).
\bibitem{Canton:2002} S. E. Canton, A. J. Yencha, E. Kukk, J. D.
Bozek, M. C. A. Lopes, G. Snell, and N. Berrah, Phys. Rev. Lett.,
{\bf 89}, 045502 (2002).
\bibitem{Majernikova:2002} E. Majern\'{\i}kov\'a, J. Riedel, and S. Shpyrko,  Phys. Rev.
B {\bf 65}, 174305 (2002).
\bibitem{Long:1958} H.C. Longuet-Higgins, U. \"Opik, and M. H. L. Pryce, Proc.
Roy. Soc. London, Ser. {\bf A244}, 1 (1958).
\bibitem{Majernikova:2003} E. Majern\'{\i}kov\'a and S. Shpyrko
J. Phys.: Cond. Matter {\bf 15}, 2137 (2003).
\bibitem{Majernikova:2006:b} E. Majern\'{\i}kov\'a and  S. Shpyrko,
Phys. Rev. E {\bf 73}, 057202 (2006).
\bibitem{Majernikova:2006:a} E. Majern\'{\i}kov\'a and  S. Shpyrko
Phys. Rev. E {\bf 73}, 066215 (2006).
\bibitem{Majernikova:2008} E. Majern\'{\i}kov\'a and S. Shpyrko,
 J. Phys. A: Math. Theor. {\bf 41}, 155102 (2008).
\bibitem{Dyson}  F. J. Dyson and M. L. Mehta, J. Math. Phys.
{\bf 4}, 701 (1963).
\bibitem{Primakoff} T. Holstein, H. Primakoff, Phys. Rev. {\bf 58}, 1098
(1949).
\bibitem{Dekker} H. Dekker, J. Phys. A: Math. Gen. {\bf 19}, L1137 (1986).
\bibitem{Overhof} H. Overhof,  J. Non-Cryst. Sol., {\bf  227-230}, 15 (1998).
 \bibitem{Shklovski} B. I. Shklovskii,  B. Shapiro, B. R. Sears, P. Lambrianides and H. B.
 Shore,  Phys. Rev. B {\bf 47}, 11487 (1993).
\bibitem{Fisher:1989} M. P. A. Fisher, P. B. Weichman, G. Grinstein and D. S. Fisher,  Phys. Rev. B {\bf 40},
546 (1989).
\bibitem{Ziegler:2005} K.Ziegler, Phys.Rev.B {\bf 72}, 075120
(2005).
\bibitem{Roth:2004} R. Roth and K. Barnett, J. Phys. B:
At. Mol. Opt. Phys. {\bf 37}, 3893 (2004).
\end{thebibliography}
\end{document}